\documentclass[10pt,letterpaper]{article}

\usepackage{graphicx}
\usepackage{url}
\usepackage{pslatex}
\usepackage{amsmath,amssymb}
\usepackage{algorithm}
\usepackage{algorithmic}
\usepackage[usenames,dvipsnames]{color}
\usepackage{fancyhdr}
\usepackage{subfig}
\usepackage{multicol}
\usepackage{multirow}
\usepackage{caption}
\usepackage{wrapfig}
\usepackage{booktabs}
\usepackage{multirow}
\usepackage{xspace}
\newcommand{\kevin}[1]{}
\newcommand{\lavanya}[1]{}

\newcommand{\TODO}[1]{}

\newif\ifSQUEEZE
\SQUEEZEtrue

\newif\ifOPTION
\OPTIONtrue

\usepackage[top=1in,bottom=1in,left=1in,right=1in]{geometry}

\newcommand{\ignore}[1]{}

\usepackage[normalem]{ulem}
\usepackage[normalem]{ulem}
\usepackage{graphicx}
\usepackage{caption}
\usepackage{multirow}
\usepackage{dblfloatfix}
\usepackage{fixltx2e}
\usepackage{setspace}
\usepackage{pifont}
\usepackage{xfrac}

\fancypagestyle{plain}{
\fancyhf{}
\chead{SAFARI Technical Report No. 2015-008 (June 8, 2015)}
\cfoot{\thepage}
}

\begin{document}

\title{\Large{Simultaneous Multi Layer Access:\\
A High Bandwidth and Low Cost 3D-Stacked Memory Interface}}

\author{
\begin{tabular}[t]{cc}
  Donghyuk Lee & Gennady Pekhimenko \\
  \texttt{donghyu1@cmu.edu} & \texttt{gpekhime@cs.cmu.edu} \\
\end{tabular} \\
\begin{tabular}[t]{ccc}
  Samira Khan & Saugata Ghose & Onur Mutlu \\
  \texttt{samirakhan@cmu.edu} & \texttt{ghose@cmu.edu} & \texttt{onur@cmu.edu} \\
	\multicolumn{3}{c}{} \\
	\multicolumn{3}{c}{Carnegie Mellon University} \\
	\multicolumn{3}{c}{} \\
	\multicolumn{3}{c}{SAFARI Technical Report No. 2015-008}
\end{tabular}
}

\date{June 8, 2015}

\maketitle

\begin{abstract}

Limited memory bandwidth is a critical bottleneck in modern systems.
3D-stacked DRAM enables higher bandwidth by leveraging wider
Through-Silicon-Via (TSV) channels, but today's systems cannot fully exploit
them due to the limited internal bandwidth of DRAM. DRAM reads a whole row
simultaneously from the cell array to a row buffer, but can transfer only a
fraction of the data from the row buffer to peripheral IO circuit, through a
limited and expensive set of wires referred to as global bitlines. In presence
of wider memory channels, the major bottleneck becomes the limited data
transfer capacity through these global bitlines. Our goal in this work is to
enable higher bandwidth in 3D-stacked DRAM without the increased cost of adding
more global bitlines. We instead exploit otherwise-idle resources, such as
global bitlines, already existing within the multiple DRAM layers by accessing
the layers simultaneously. Our architecture, Simultaneous Multi Layer Access
(SMLA), provides higher bandwidth by aggregating the internal bandwidth of
multiple layers and transferring the available data at a higher IO frequency.

To implement SMLA, simultaneous data transfer from multiple layers through the
same IO TSVs requires coordination between layers to avoid channel conflict. We
first study coordination by static partitioning, which we call Dedicated-IO,
that assigns groups of TSVs to each layer. We then provide a simple, yet
sophisticated mechanism, called Cascaded-IO, which enables simultaneous access
to each layer by time-multiplexing the IOs, while also reducing design effort.
By operating at a frequency proportional to the number of layers, SMLA provides
a higher bandwidth (4X for a four-layer stacked DRAM). Our evaluations show
that SMLA provides significant performance improvement and energy reduction
(55\%/18\% on average for multi-programmed workloads, respectively) over a
baseline 3D-stacked DRAM with very low area overhead.

\end{abstract}

\section{Introduction}

Main memory, predominantly built using DRAM, is a critical performance
bottleneck in modern systems due to its limited
bandwidth~\cite{burger-sigarch96,solihin-isca10}. With increasing core count
and memory intensive applications, memory bandwidth is expected to become a
greater constraint in the future~\cite{dean13}. For the last few decades, DRAM
vendors provided higher bandwidth by using higher IO frequencies, increasing
the bandwidth available per pin (a 17x improvement over the last decade, from
$400 Mbps$ in DDR2 to $7 Gbps$ in GDDR5~\cite{smith12}). However, further
increase in frequency is challenging due to higher energy consumption and
logical complexity. Recent development in 3D integration through
\emph{Through-Silicon Vias} (TSVs) provides an alternate way to provide higher
bandwidth. TSVs enable wider IO interface among vertically stacked layers in
3D-stacked DRAM
architectures~\cite{wideio,jedec-wideio2,hmc,kang-isscc09,jedec-hbm,lee-isscc14}.

Even though the TSV technology enables higher data transfer capacity, existing
3D-stacked DRAMs cannot fully take advantage of the potential wide interface.
For example, Wide-IO~\cite{wideio} offers an extremely high bus width (512 bit,
$16-64$ times wider than conventional DRAM chips), but can only operate at much
lower frequencies ($200-266MHz$) than conventional DRAMs (e.g., as high as
$2133MHz$ for DDR3). As a result, the effective bandwidth increase is an order
of magnitude lower than the bus width increase. Based on our analysis of the
DRAM hierarchy (explained in Section~\ref{sec:background}), 3D-stacked DRAMs
cannot fully exploit the wider IO interface due to their limited internal
bandwidth. DRAM reads a whole row simultaneously from the cell array to a row
buffer, but can transfer only a fraction of data in the row buffer through a
limited set of wires (global bitlines) and sense amplifiers. Therefore, the
number of global bitlines dominates the DRAM internal data transfer capacity.
One naive way to enable higher bandwidth is to add more global bitlines and
sense amplifiers~\cite{jedec-wideio2,jedec-hbm}. However, the area and energy
constraints make this intuitive and simple solution extremely
expensive.\footnote{We estimate that global sense amplifiers (and their
corresponding control logic) consume 5.18\% of the total area in a 55nm DDR3
chip~\cite{vogelsang-micro10,rambus-power10}. Considering that 3D-stacked DRAM
already contains many more global sense amplifiers (see
Section~\ref{sec:bandwidth}), adding further sense amplifiers may be quite
costly. The sense amplifier also consumes a large amount of energy because it
is typically implemented as a differential amplifier, whose performance is
strongly dependent on standby current~\cite{dram-circuit-design,razavi}.}

Our goal in this work is to enable higher bandwidth in 3D-stacked DRAM without
effectively increasing the cost by adding extra global sense amplifiers. In
order to design such a system, we make the observation that a large number of
global bitline interfaces already exist across the multiple layers of a
3D-stacked DRAM, but only a fraction of them are being enabled at any
particular point of time. We exploit these otherwise-idle interfaces to access
multiple DRAM layers \emph{simultaneously}, which can increase the data
delivery bandwidth to the TSV-based IO interface (which already has enough
bandwidth to serve this additional data). We call this architecture {\em
Simultaneous Multi Layer Access} ({\em SMLA}). Using SMLA, multiple global
bitline interfaces in multiple layers provide data to IO, enabling the IO
interface (that is vertically connected across all stacked layers) with enough
data to warrant transfers at a much higher frequency than conventional
3D-stacked DRAM.

\begin{figure}[h]
	\center
	\vspace{-0.1in}
	\subfloat[Baseline] {
		\includegraphics[height=1.8in]{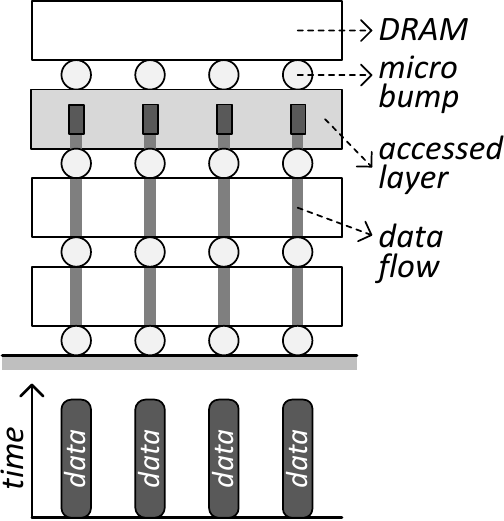}
		\label{fig:baseline}
	}
	\subfloat[Dedicated-IO] {
		\includegraphics[height=1.8in]{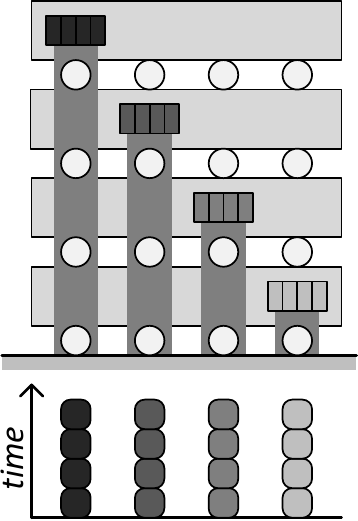}
		\label{fig:dedicated}
	}
	\subfloat[Cascaded-IO] {
		\includegraphics[height=1.8in]{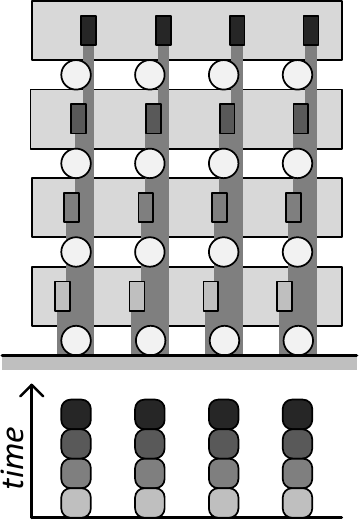}
		\label{fig:cascaded}
	}
	\vspace{-0.05in}
	\caption{Single Layer (baseline) vs. Multi Layer Access}
	\label{fig:summary}
	\vspace{-0.1in}
\end{figure}

To implement the idea of SMLA, simultaneous data transfer from multiple layers
through the same IOs requires co-ordination between layers to avoid channel
contention. One simple way to enable an SMLA architecture is to restrict the
sharing of channels and assign IOs dedicated to each layer. We refer to this
simple solution as {\em Dedicated-IO}, where fractions of TSVs form an IO group
that is dedicated to a single layer (Figure~\ref{fig:dedicated}). Each layer
has access to a smaller number of IOs, but can transfer the same amount of data
by operating its IOs at a higher frequency. As each layer transfers data
simultaneously at a higher frequency, Dedicated-IO enables bandwidth
proportional to the number of layers (4X for a four-layer stacked DRAM) than
the baseline system (Figure~\ref{fig:baseline}).\footnote{A new design from
Hynix~\cite{jedec-hbm} proposes to use exclusive IO interfaces for each layer
of 3D-stacked DRAM, but enables higher bandwidth by increasing the internal
bandwidth with extra global bitlines. Dedicated-IO, on the contrary, enables
higher bandwidth at low-cost by leveraging existing global bitline interfaces
and operating at a higher frequency (described in Section~\ref{related}).}
While this architecture enables higher bandwidth, it has two disadvantages.
First, as each layer requires dedicated connection to specific channels, the
design of each layer is not uniform anymore, resulting in higher manufacturing
cost. Second, the IO clock frequency scales linearly with the number of layers,
resulting in greater dynamic energy consumption.

To solve these problems, we propose {\em Cascaded-IO}, which exploits the
architectural organization of TSV-based interface, where an upper layer
transfers its data through lower layers. Cascaded-IO
(Figure~\ref{fig:cascaded}) enables simultaneous access to each layer by
time-multiplexing the IOs. In this design, each layer first sends its own data
and then sends data transferred from the upper layers. By operating at a
frequency proportional to the number of layers, Cascaded-IO provides a much
higher bandwidth (4X for a four-layer stacked DRAM). While Cascaded-IO operates
at a higher frequency, we observe that only the bottom layer theoretically
needs to operate at the highest frequency, as this is the only layer that
transfers data from all upper layers. We propose to reduce the frequency of
other layers and optimize the frequency individually for each layer based on
the bandwidth requirements. {\color{black} As a result, Cascaded-IO enables
higher bandwidth and lower energy consumption with a homogeneous DRAM design.
While each layer in Cascaded-IO can operate at different frequency, we still
expect that Cascaded-IO requires similar validation effort to existing DRAMs,
as existing DRAMs guarantee correct operation at slower timing/frequency than
their specification.}

Our work makes the following contributions:

\begin{enumerate}

	\item We propose a new 3D-stacked DRAM organization Simultaneous Multi Layer
	Access (SMLA), which enables higher bandwidth with low cost by leveraging the
	otherwise-idle interfaces in multiple layers of 3D-stacked memory.

	\item We introduce mechanisms to transfer data from multiple layers without
	conflicting in the shared IO interface of 3D-stacked DRAMs. Simply assigning
	IOs to different layers (Dedicated-IO) avoids channel contention, but
	increases manufacturing cost and energy consumption. We propose an elegant
	and simple mechanism called Cascaded-IO, which time-multiplexes shared IOs so
	that each layer in 3D-stacked DRAM first transfers data from own layer and
	then from upper layers. Our mechanism has three major benefits. First, it
	provides bandwidth proportional to the number of layers in 3D-stacked DRAM
	with low cost. Second, it avoids non-uniformity in the design by
	time-multiplexing the IO interface, reducing manufacturing cost. Third, it
	enables higher bandwidth with lower energy consumption by optimizing IO
	frequency based on each layer's bandwidth requirement.

	\item Our extensive evaluation of 3D-stacked DRAMs on 31 applications from
	SPEC CPU2006, TPC, and STREAM applications suites shows that our proposed
	mechanisms significantly improve performance and reduce energy consumption
	(55\%/18\% on average of 16-core multi-programmed workloads, respectively)
	over conventional 3D-stacked DRAMs (with no support for SMLA).

\end{enumerate}

\section{3D-Stacked DRAM Bandwidth Constraints} \label{sec:background}

To understand the internal bandwidth bottleneck of 3D-stacked DRAM, we first
describe the unique design of the IO interface in 3D-stacked DRAM
(Section~\ref{sec:3d-stacked}). Then, we walk through the DRAM architectural
components that make up the datapath used for both read and write requests. We
analyze the bandwidth of each component along this datapath in
Section~\ref{sec:dram}, and study a trade-off between bandwidth and area in
Section~\ref{sec:bandwidth}.

\subsection{Using TSVs to Increase IO Bandwidth}
\label{sec:3d-stacked}

\begin{figure}[b]
	\vspace{-0.1in}
	\center
	\subfloat[Overall] {
		\includegraphics[height=1.2in]{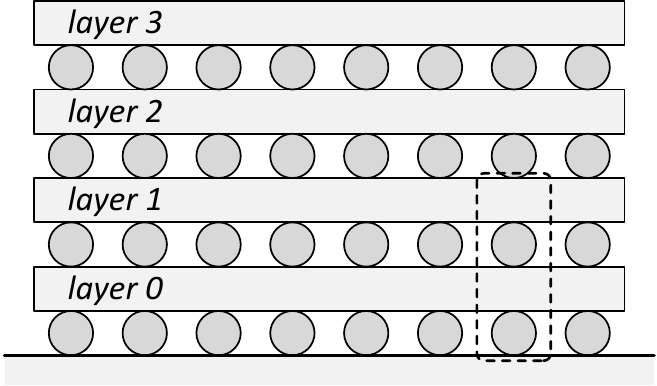}
		\label{fig:3d_stacked}
	} \hspace{0.4in}
	\subfloat[Detailed] {
		\includegraphics[height=1.2in]{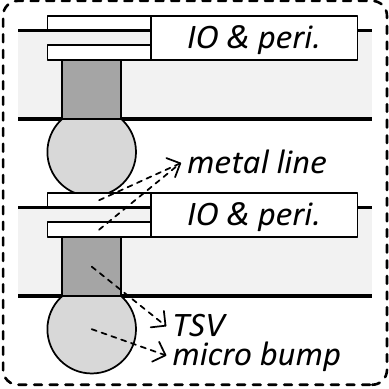}
		\label{fig:bump_tsv}
	}
	\vspace{-0.05in}
	\caption{TSV Interface in 3D-Stacked DRAM}
	\label{fig:3d_dram}
\end{figure}

We first explain the detailed organization of a 3D-stacked DRAM that integrates
a much wider IO bus than conventional DRAM. Figure~\ref{fig:3d_stacked} shows a
3D-stacked DRAM consisting of four layers that are connected over micro-bump
and TSV interfaces. A TSV interface vertically connects all of the layers. The
bottom of the four-stacked layer is either directly placed on top of a
processor chip or connected to a processor chip by metal wires.

Figure~\ref{fig:bump_tsv} details the organization of these TSV and micro-bump
connections. Two metal lines connect each layer to these interfaces. The top
metal line is connected to the micro-bump of its upper layer, and the bottom
metal line is connected to the TSV. At the bottom of the chip, this TSV is
exposed and connected to a micro-bump, connecting to the top metal line of the
lower layer. These two metal lines are eventually connected by via or over
peripheral circuits. In this way, several such layers can be stacked one on top
of another.

TSVs provide two major benefits to increase memory bandwidth and energy
efficiency. First, due to the small feature size of modern TSV technologies
($10-35\mu$m pitch for
TSVs~\cite{wideio,west12,huygherbaert10,harvard-mwscas11} vs.\ $90\mu$m pitch
for conventional pads), 3D-stacked DRAM can integrate hundreds of TSVs for its
connections between layers. Second, the small capacitance of a TSV reduces
energy consumption. Compared to conventional DRAM, whose off-chip IO are
connected by long bonding wires and metal connections in the package, TSV
connections between layers are very short, leading to lower capacitance and
energy consumption during data transfer. As a result, 3D-stacked DRAM offers a
promising DRAM architecture that can enable both higher bandwidth and energy
efficiency.

\subsection{DRAM Bandwidth Analysis} \label{sec:dram}

\begin{figure}[ht]
	\vspace{-0.1in}
	\center
	\subfloat[Organization] {
		\includegraphics[height=1.7in]{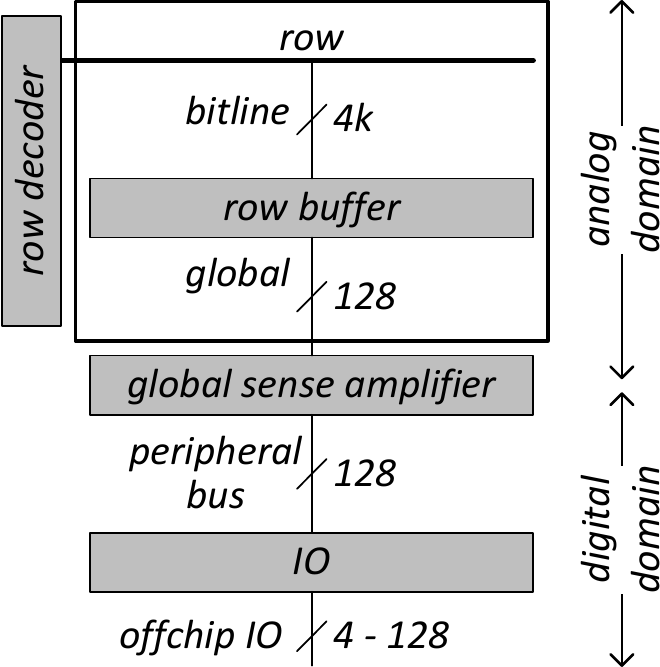}
		\label{fig:dram_io}
	}
	\hspace{0.25in}
	\subfloat[Timing] {
		\includegraphics[height=1.7in]{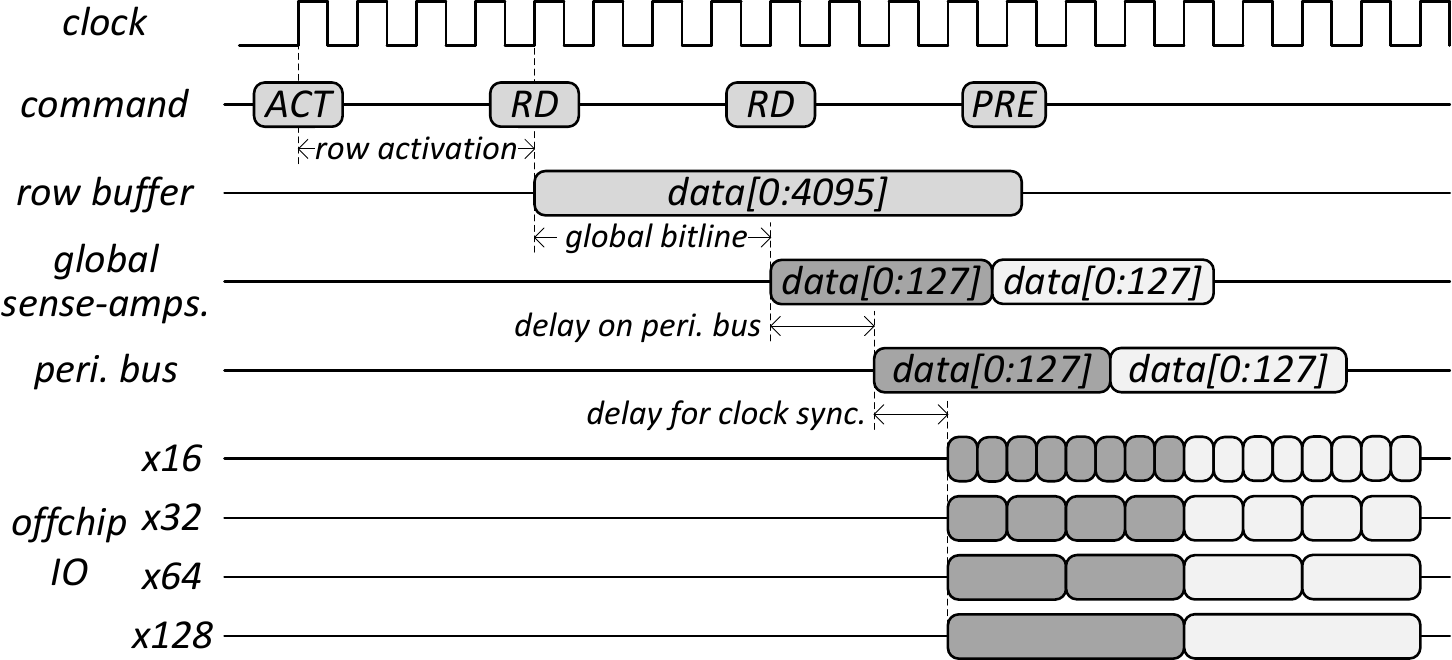}
		\label{fig:dram_timing}
	}
	\vspace{-0.05in}
	\caption{Overview of DRAM Organization}
	\vspace{-0.1in}
	\label{fig:dram}
\end{figure}

DRAM has two kinds of accesses (read and write) that transfer data through
mostly the same path between its cells and off-chip IO.
Figure~\ref{fig:dram_io} shows the read datapath from the DRAM cells to IO.
When activating a row, all data in a row is transferred to a row buffer (a row
of local sense amplifiers) through bitlines. Then, when issuing a read command
with a column address, small fraction of data in the row buffer (corresponding
to the issued column address) is read by the global sense amplifiers.
Peripheral logic then transfers this data to the off-chip IO. To write data,
commands containing data are issued, and peripheral logic transfers this data
to the global sense amplifiers that write the data to a row buffer, which then
writes the data to the cells corresponding to the requested address.

At each step, the individual structures along the datapath have their own data
transfer rate (bandwidth). For a read access, we explore the bandwidth of each
step in detail. When activating a row, all data within the row (e.g., $8
Kbits$~\cite{micron_spec}) is transferred to the row buffer, which takes about
$13 ns$ (based on {\tt tRCD} timing constraint~\cite{micron_spec}).  Therefore,
the bandwidth of the activation step is about $78.8 GBps$
(Giga-Bytes-per-Second). After migrating data to a row buffer, global sense
amplifiers read $64-128$ bits of data from the row buffer for about $3-5 ns$
through the global bitline.\footnote{Due to the high latency of global bitline,
DRAM operates at much lower frequency ($200-266MHz$) than its IO
frequency~\cite{smith12}. Therefore, the ratio between IO frequency and in-DRAM
frequency is same as the ratio between global bitline width and IO width,
referred to as {\em Prefetch Size} in DRAM.} Therefore, the bandwidth of the
global biltine is about $2-4 GBps$. The data read by global sense amplifiers is
transfered to the IO interface, which can operate at frequencies as high as $2
GHz$. Considering that the off-chip IO bus width can range from 4 to 128 bits,
the off-chip bandwidth can be in the range of $1-32 GBps$.

Based on the above analysis, in conventional DRAMs that have small off-chip bus
width (4 to 32 bits), the overall DRAM bandwidth can be bottlenecked by the
bandwidth of the off-chip IO interface. However, with a wide-width off-chip IO
interface (e.g., the 512-bit TSV interface on 3D-stacked DRAM), the global
sense amplifiers turn into the dominant bottleneck of the overall DRAM
bandwidth. Unfortunately, increasing the bandwidth of the global bitline
interface requires large area overhead due to the need for additional global
sense amplifiers and global biltines (as we will discuss in
Section~\ref{sec:bandwidth}). Furthermore, increasing the bandwidth of the
global bitline interface requires a significant amount of additional energy. A
differential amplifier, a popular primitive to implement the global sense
amplifier, consumes significant levels of standby current. Due to the limited
available global sense amplifier bandwidth, higher-frequency TSV interfaces for
3D-stacked DRAMs are currently unnecessary.

To show the relationship between the operating frequency and the bus width of
the IO interface, we draw timing diagrams for the data bus during a read access
(Figure~\ref{fig:dram_timing}). We divide the DRAM hierarchy into two domains,
{\em (i)} an analog domain, whose components consist of sensing structures and
the data repository (the row buffer and global sense amplifiers in
Figure~\ref{fig:dram_io}), and {\em (ii)} a digital domain, whose components
(the peripheral bus and off-chip IO in Figure~\ref{fig:dram_io}) transfer data
with normal peripheral transistors and full swing of the data value (0 or 1).

In the analog domain, a data repository directly maps to a sensing structure
with a fixed metal wire, and accessing this repository (the row buffer) from a
global sense amplifier takes a fixed latency. During this global bitline
latency, only one bit of data can be transferred over one global bitline.
Therefore, the bandwidth of the global bitline interface is tightly coupled
with the number of global bitlines and global sense amplifiers (128 bits in
this example). Considering that each global sense amplifier is directly
connected to a row buffer through a single wire and detects the intermediate
voltage value (between 0 to {\tt Vdd}) of a row buffer, it is difficult to
divide the global bitline interface into multiple stages to get the benefits of
pipelining.

On the other hand, the peripheral interface (from the global sense amplifiers
to off-chip IO) and IO interface (toward off-chip) can be divided into multiple
stages and implemented in a pipelined manner. Therefore, as shown in
Figure~\ref{fig:dram_timing}, the 128 bits of data from global sense amplifiers
can be transferred through either a narrow IO bus with higher frequency (e.g.,
DDR3 with a 16-bit bus) or wide IO bus with lower frequency (e.g., Wide-IO with
a 128-bit bus per channel).~\footnote{Conventional 3D-stacked DRAM consists of
multiple channels that can operate independently -- i.e, four channels of
Wide-IO~\cite{wideio,jedec-wideio}, four or eight channels of
Wide-IO2~\cite{jedec-wideio2} and HBM~\cite{jedec-hbm}.} In the ideal case,
assuming that the global biltine interface provides enough data to the
peripheral interface, the wide bus in off-chip IO can transfer the data at
higher frequency, enabling much higher bandwidth (i.e., $128 bits \times 2GHz =
32GBps$).

Based on this analysis, we observe that even though 3D-stacked DRAM can
potentially enable much greater bandwidth, the limited bandwidth on the global
bitline interface significantly limits the actually available bandwidth.

\subsection{Bandwidth vs. Global Sense Amplifiers} \label{sec:bandwidth}

\begin{figure}[ht]
	\centering
	\includegraphics[width=0.65\linewidth]{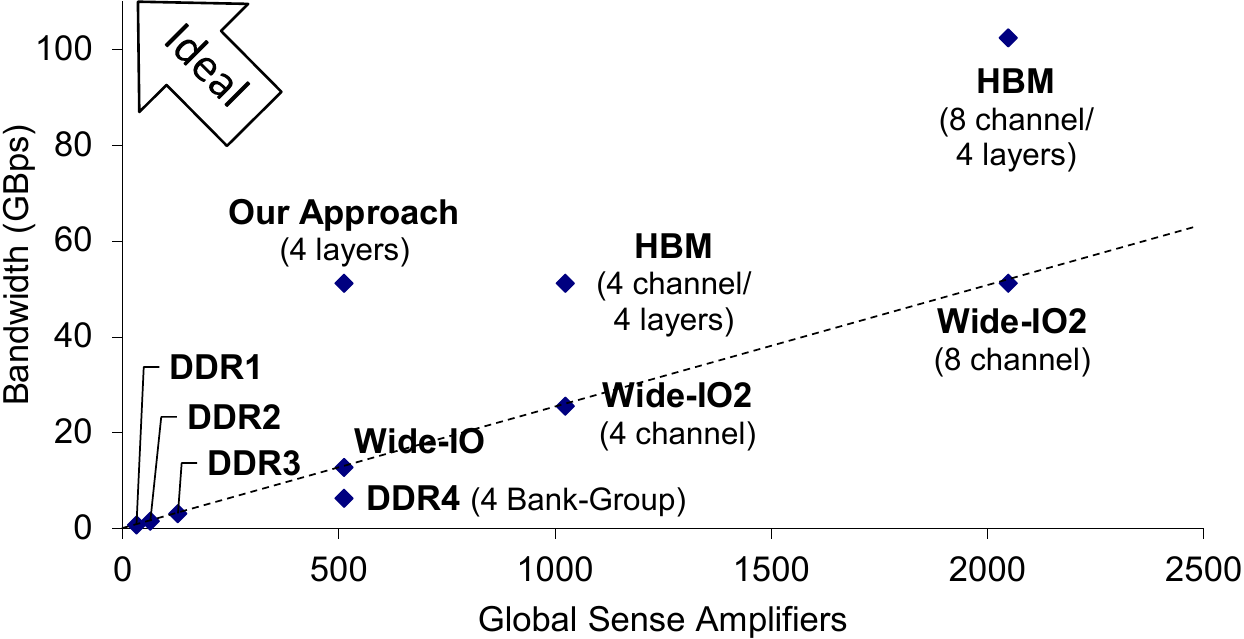}
	\vspace{-0.05in}
	\caption{Bandwidth vs. Global Sense Amplifiers (Selecting DRAM bins that
	in-DRAM frequency is 200MHz)}
	\vspace{-0.1in}
	\label{fig:trend_bandwidth}
\end{figure}

Figure~\ref{fig:trend_bandwidth} shows the relationship between the bandwidth
of DRAM-based memory systems (one chip for conventional DRAM, four layers for
3D-stacked DRAM) and the number of global sense amplifiers in a DRAM chip.  We
conservatively plot the minimum number of global sense amplifiers that achieves
the proposed bandwidth in each design.~\footnote{For example, for DDR3 DRAM
with an x16 IO bus and 8 times more global bitlines (x128) than IO bus, we
assume that DDR3 has 128 global sense amplifiers (and 128 global bitlines),
even though DDR3 can be designed to have multiple sets of global bitline
interfaces, as this depends on the process technology and product design.}

Figure~\ref{fig:trend_bandwidth} shows that the bandwidth of DRAM mostly has
been a linear function of the number of global sense amplifiers across several
generations. In early DRAMs,  the number of global sense amplifier structures
was small (e.g., 64 for a DDR2 chip) and most likely was not the dominant
source of DRAM area. Therefore, DRAM vendors had scaled the number of global
sense amplifiers with IO frequency to increase the bandwidth.  However,
recently proposed 3D-stacked DRAMs~\cite{jedec-wideio,jedec-wideio2,jedec-hbm}
drastically increase the number of global sense amplifiers. For example,
HBM~\cite{jedec-hbm,lee-isscc14} offers the highest bandwidth, but requires a
large number of global sense amplifiers (16X more than DDR3). This drastic
increase comes from introducing TSV technologies that enable a wider,
area-efficient bus. In the presence of the wider bus, the dominant bottleneck
in achieving higher bandwidth now has shifted to the efficient implementation
of global sense amplifiers.

We estimate the area of global sense amplifiers (and their corresponding
control logic) to be 5.18\% of the total chip area in a 55nm DDR3
DRAM~\cite{vogelsang-micro10,rambus-power10}. Taking into account {\em (i)} the
large area of global sense amplifiers in DDR3 and {\em (ii)} the recent
increase in the number of sense-amplifiers in 3D-stacked DRAMs, we expect that
adding more global sense amplifiers will become extremely expensive. As we
explain in Section~\ref{sec:motivation}, in this paper, we enable much higher
bandwidth at low-cost using the existing structure of 3D-stacked DRAM,
approaching the ideal case (the top left corner of
Figure~\ref{fig:trend_bandwidth}).

\section{Opportunities for Increasing Bandwidth}
\label{sec:motivation}
\label{sec:compare}

Even though 3D-stacked DRAM integrates a much wider IO bus through the use of
TSVs, the limited internal bandwidth within the DRAM chip still constrains the
total bandwidth that we can extract from this wider IO, as we analyzed in
Section~\ref{sec:dram}. While one solution may be to expand the number of
global bitlines and sense amplifiers, this can be a costly prospect (as we show
in Section~\ref{sec:bandwidth}). Instead, we would like to investigate
orthogonal approaches to overcome the global bitline bottleneck, that exploit
the existing structures more efficiently.

\begin{figure}[ht]
	\begin{minipage}[ht]{0.49\linewidth}
	\vspace{-0.1in}
		\center
		\subfloat[DRAM Module] {
			\includegraphics[height=1.20in]{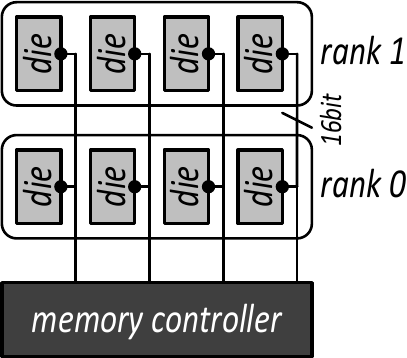}
			\label{fig:conventional}
		}
		\subfloat[3D-Stakced] {
			\includegraphics[height=1.20in]{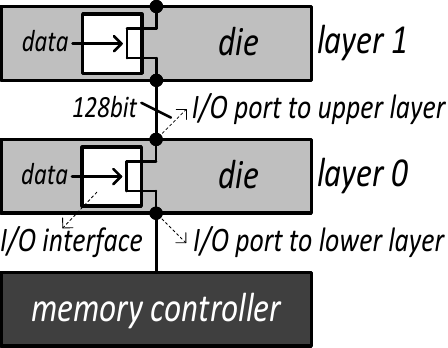}
			\label{fig:3D}
		}
		\vspace{-0.05in}
		\caption{DRAM Module vs. 3D-DRAM} \label{fig:compare}
		\vspace{-0.1in}
	\end{minipage}
	\hspace{0.1in}
	\begin{minipage}[ht]{0.49\linewidth}
	\vspace{-0.1in}
		\center
		\subfloat[Dedicated-IO] {
			\includegraphics[height=1.20in]{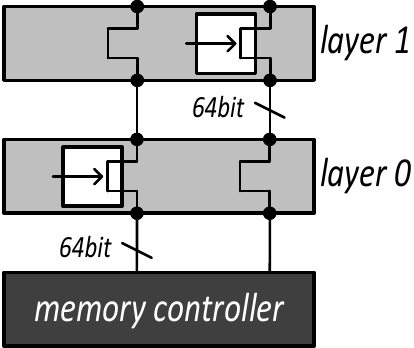}
			\label{fig:multirank_dio}
		}
		\subfloat[Cascaded-IO] {
			\includegraphics[height=1.20in]{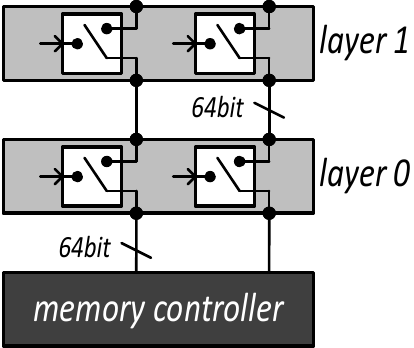}
			\label{fig:multirank_cio}
		}
		\vspace{-0.05in}
		\caption{Proposals for More Bandwidth} \label{fig:approach}
		\vspace{-0.1in}
	\end{minipage}
\end{figure}

We examine the architectural similarities and differences between 3D-stacked
DRAM and conventional DRAM modules, and use this comparison to identify
challenges and opportunities that are unique to the 3D-stacked DRAM design. We
will use these observations to drive our low-cost approach. To aid with this
comparison, we provide a side-by-side logical view of these two architectures
in Figure~\ref{fig:compare}, and focus on three major observations:

\begin{enumerate}

\item {\em Each layer in 3D-stacked DRAM acts as a rank, operating
independently but sharing the IO interface with each other, which is similar to
the rank organization of conventional DRAM modules.}

Figure~\ref{fig:conventional} shows a DRAM module that has two ranks, each
consists of four DRAM chips. All DRAM chips in a rank share control logic and
operate in lockstep. The two ranks share the IO interfaces such that requests
to different ranks are served serially. Figure~\ref{fig:3D} shows the
organization of a 3D-stacked DRAM-based system that has two stacked layers.
Similar to the multi-rank DRAM module, each layer in 3D-stacked DRAM can
operate independently, but still shares the IO interface (TSV connection).
Therefore, the TSV interface transfers data from different layers serially.

\item {\em Only one stacked chip forms a rank in 3D-stacked DRAM, compared to
multiple chips in conventional DRAM module.}

Figure~\ref{fig:compare} shows that four conventional DRAM chips form a rank by
dividing the 64-bit IO bus into four smaller 16-bit IO buses, with each smaller
bus connected to a different chip. In contrast, each 3D-stacked DRAM layer is
made up of a single chip, with the entire TSV bus connected to it.

\item {\em Each layer of 3D-stacked DRAM has two ports that can be easily
decoupled by peripheral logic, whereas conventional DRAM has only a single
port.}

As shown in Figure~\ref{fig:bump_tsv}, each stack layer in 3D-stacked DRAM is
connected to its neighboring upper stack layer through the topmost metal layer.
Each stack layer is also connected to its lower stack layer through one of its
middle metal layers (different from the topmost metal layer).  These two metal
layers can be connected within the DRAM stack layer by either fixed via or
peripheral logic. These two different metal layers form two independent IO
ports, whose connectivity can be controlled by peripheral logic. In the general
organization of 3D-stacked DRAM, the two metal wires that form the DRAM input
ports are simply connected by via structures, meaning that they are connected
but can be easily decoupled to support independent data transfers on each port.

\end{enumerate}

As we described above, the major bandwidth bottleneck in 3D-stacked DRAM-based
memory system is the global bitline interface in DRAM.  Our approach to solve
this problem is inspired by the organization of the conventional DRAM module,
where multiple DRAM chips operate simultaneously, thereby increasing the
overall bandwidth of the memory system (Figure~\ref{fig:conventional}). From
our observations, we conclude that while the 3D-stacked layers individually
look different from the conventional DRAM architecture perspective, the ability
to decouple input ports can allow us to treat multiple layers similarly to the
way we treat multiple chips within DRAM.

\section{Simultaneous Multi Layer Access} \label{sec:smla}

We propose {\em Simultaneous Multi Layer Access} (SMLA), a new mechanism that
enables the concurrent use of multiple stack layers (chips) within 3D-stacked
DRAM. SMLA exploits most of the existing DRAM structures within the chip,
altering the IO logic to include multiplexing that supports these concurrent
operations. This allows the bandwidth gains obtained by SMLA to be
complementary to the gains that could be otherwise obtained from scaling these
DRAM structures (e.g., adding more global bitlines and sense amplifiers).

In this section, we will describe two implementations for coordinating these
multiplexed IOs. The first, \emph{Dedicated-IO}, divides the TSV bus into
several groups, with each group being dedicated to the IO of a single stack
layer. The second, \emph{Cascaded-IO}, explores the idea of time-multiplexing
the bus instead of partitioning it, by exploiting our observation that we can
decouple the input ports to in effect vertically segment the TSVs
(Section~\ref{sec:compare}). As we will discuss in Section~\ref{sec:cio}, the
Cascaded-IO implementation allows the circuit design of each stack layer to be
identical, reducing design effort, and offers a more energy-efficient approach
to implement SMLA.

\subsection{Dedicated-IO} \label{sec:dio}

The Dedicated-IO implementation divides the wide TSV bus into several narrower
groups, each of which is dedicated to a single stack layer (a DRAM chip). As
Figure~\ref{fig:multirank_dio} shows, in the case of a two layer 3D-stacked
DRAM, each layer has its own IO connection, whose width is half of the total IO
width $W$ (128 bits in the figure). With $L$ layers working simultaneously,
since each still contains $W$ global bitlines and sense amplifiers, we can now
transmit $W \times L$ bits to the memory controller in the time that it
previously took to transmit $W$ bits, by increasing the IO clock from its
original frequency $F$ to a new frequency $F \times L$.\footnote{Naturally, the
magnitude of $L$ is limited, as the frequency $F \times L$ cannot exceed the
total attainable frequency for the peripheral interface and IO interface.} For
our two-layer example in Figure~\ref{fig:multirank_dio}, we can increase the IO
frequency, and therefore the bandwidth, to twice that of the baseline
3D-stacked DRAM (Figure~\ref{fig:3D}) by having each layer send two sets of
64-bit data at $2F$ frequency.

\subsubsection{Implementation Details} \label{sec:dio_detail}

\begin{figure}[ht]
	\hspace{0.2in}
	\vspace{-0.2in}
	\center
	\subfloat[Baseline] {
		\includegraphics[height=1.1in]{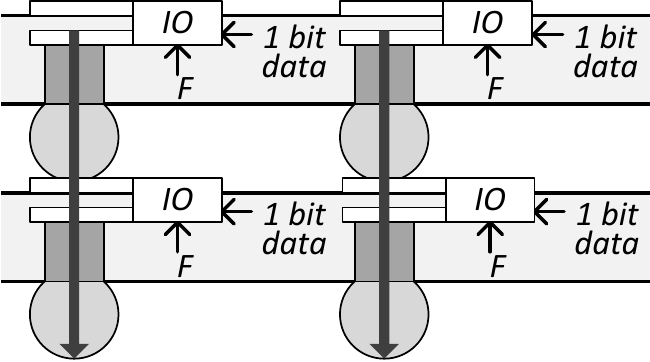}
		\label{fig:sio_detail}
	} \hspace{0.5in}
	\subfloat[Dedicated-IO] {
		\includegraphics[height=1.1in]{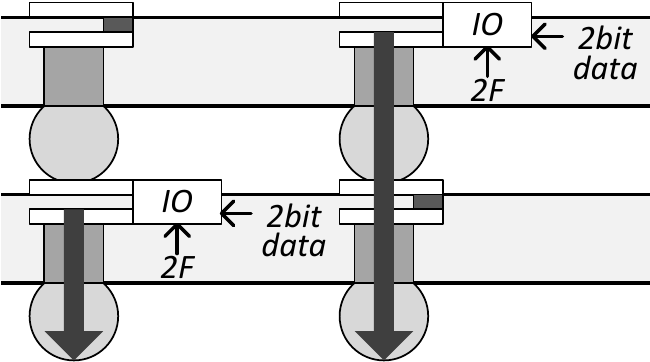}
		\label{fig:dio_detail}
	}
	\vspace{-0.05in}
	\caption{Simultaneous Multi Layer Access with Dedicated-IO}
	\label{fig:dio}
\end{figure}

Figure~\ref{fig:dio} compares the detailed organizations of existing 3D-stacked
DRAM (Figure~\ref{fig:sio_detail}), where all layers share all TSVs, and a
3D-stacked DRAM using Dedicated-IO (Figure~\ref{fig:dio_detail}). To clearly
explain the access scenarios for both cases, we use a simplified organization
of 3D-stacked DRAM, consisting of two layers, and we focus only on a two-bit
subset of each layer (i.e., it contains two TSVs as IO for data transfer, and
each layer has two bitlines that can deliver two bits every clock period, which
is $\sfrac{1}{F}$ for the baseline).  In this example, the Dedicated-IO-based
3D-stacked DRAM dedicates the left TSV to the lower stack layer, and the right
TSV to the upper layer.

In the case of the baseline 3D-stacked DRAM, only one selected layer (the upper
layer in this example) can transfer data during a single clock period.
Therefore, the upper layer transfers one bit per TSV (two bits in total).  On
the other hand, for the Dedicated-IO-based 3D-stacked DRAM, both layers are
transferring data through the dedicated TSV of each layer. Note that it is
still possible to increase the IO frequency of 3D-stacked DRAM, as conventional
DRAMs have a much higher IO frequency (see Section~\ref{sec:background}).
Therefore, the Dedicated-IO-based 3D-stacked DRAM can transfer two bits of data
\emph{from each layer} during one baseline clock period ($\sfrac{1}{F}$) by
doubling the IO frequency.  As a result, Dedicated-IO enables twice the
bandwidth of the conventional 3D-stacked DRAM by {\em (i)} simultaneously
accessing both layers (doubling the available data per TSV) and {\em (ii)}
transferring data at doubled frequency.

\subsubsection{Implementation Overhead} \label{sec:dio_overhead}

While Dedicated-IO enables much higher bandwidth, there are several drawbacks
in this implementation on a real memory system. First, each layer of the
3D-stacked DRAM has physically different connections. Therefore, the
manufacturing cost of the Dedicated-IO-based 3D-stacked DRAM may be higher than
conventional 3D-stacked DRAM. Second, with an increasing number of layers, the
IO clock frequency scales linearly, resulting in higher dynamic energy
consumption.

\subsection{Cascaded-IO} \label{sec:cio}

Cascaded-IO reorganizes the IO interface in order to address the design
shortcomings of Dedicated-IO, through the use of time-sliced multiplexing. As
Figure~\ref{fig:multirank_cio} shows, the IO interface of each layer has a
multiplexer that can transfer either $W$ bits of its own data, or $W$ bits of
data from an upper layer. By performing these transfers at $F \times L$
frequency, each layer can send $W$ data over the baseline clock period
$\sfrac{1}{F}$ without having to partition the data (i.e., each layer gets to
transmit its data every $L$ clock cycles). This mechanism is only possible in a
3D-stacked DRAM that contains two independent ports connected by peripheral
circuits, as we discussed in Section~\ref{sec:compare}.

\subsubsection{Implementation Details} \label{sec:cio_detail}

\begin{figure}[ht]
	\center
	\vspace{-0.1in}
	\subfloat[Vertical Structure] {
		\includegraphics[height=1.6in]{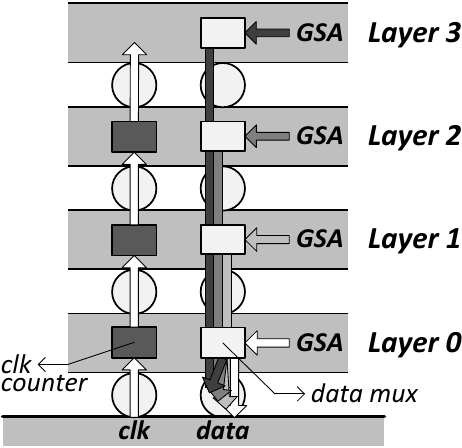}
		\label{fig:cio_org}
	}
	\subfloat[Identical Clock Frequency] {
		\includegraphics[height=1.6in]{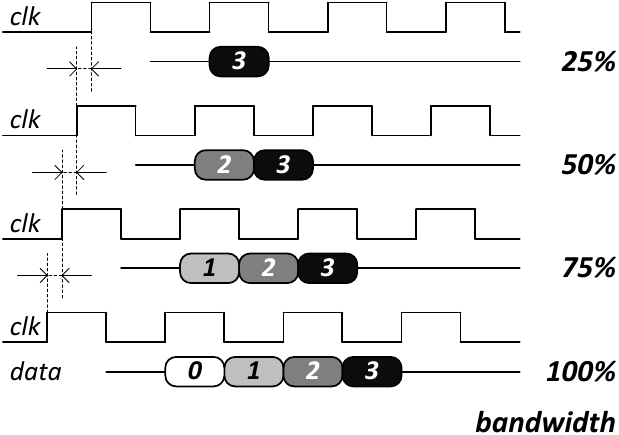}
		\label{fig:cio_sclk}
	}
	\subfloat[Optimized Clock Frequency] {
		\includegraphics[height=1.6in]{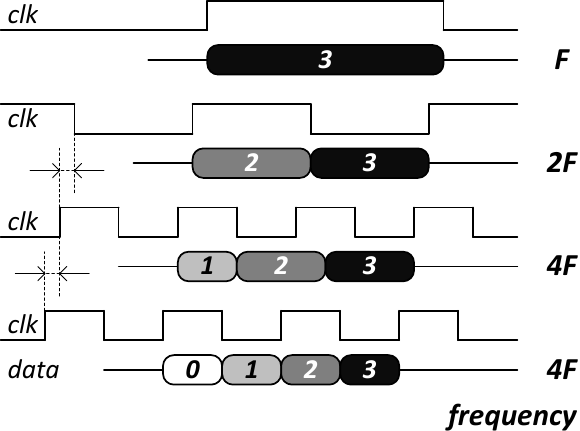}
		\label{fig:cio_rclk}
	}
	\vspace{-0.05in}
	\caption{Simultaneous Multi Layer Access with Cascaded-IO} \label{fig:cio}
	\vspace{-0.1in}
\end{figure}

Figure~\ref{fig:cio_org} shows the vertical structure of Cascaded-IO on a
four-layer 3D-stacked DRAM. We again turn to a simplified example to illustrate
the operation sequence, this time focusing on a one-bit subset of each layer,
and with all bits connected to the same TSV. In this setup, the clock
frequency, with respect to the baseline DRAM clock frequency $F$, will be $4F$.
The data interface in each layer contains a multiplexer that selects from one
of two paths: {\em (i)} a connection to the upper stack layers to transmit its
bit of data, and {\em (ii)} a data wire from the layer's global sense
amplifiers, connected to its own cell.

In a Cascaded-IO-based memory system, each layer will first fetch its own data
from its cell. The multiplexer will then drive this data as the first data
transfer, which will take the data from a layer and send it to the next layer
below. Over the next three cycles, the layer will send data that it received
from the layer above it. In essence, this will pipeline the data of all four
layers over the four cycles, as shown in Figure~\ref{fig:cio_sclk}.  Layer~1,
for example, will drive its own data in the first cycle, sending it to Layer~0.
In the second cycle, Layer~1 will receive Layer~2's data, and will now send
that down to Layer~0. During this second cycle, Layer~2 will receive the data
from Layer~3, sending that down to Layer~1 so that in cycle three, Layer~1 can
now send the Layer~3 data down to Layer~0.

One side effect of this pipelined approach is that not all layers contain
useful data to send to its lower level every cycle. In our example, Layer~1 has
nothing to send in the fourth cycle, and Layer~3, with no layers above it, only
sent useful data in the first of the four cycles. As we show in
Figure~\ref{fig:cio_sclk}, the layer-by-layer TSV bandwidth utilization grows
from 25\% at Layer~3 up to 100\% at Layer~0. Since the bandwidth in these upper
layers is going to be unused, it is wasteful to still clock them at $4F$
frequency. We exploit this tiered utilization to lower the energy utilization
of higher layers, by simply running these layers at lower frequencies. We show
the detailed effects of this energy reduction in Section~\ref{sec:energy}.

While phase-locked loops (PLLs~\cite{novof-jssc95}) are a conventional way to
generate clocks of arbitrary frequency, they consume large amounts of energy,
making them an infeasible option for our heterogeneous clock frequency domains.
Instead, Cascaded-IO adopts simple {\em two-bit clock counters}, which, when
enabled, can generate a half-frequency clock. Figure~\ref{fig:cio_org} also
shows the organization of the clock path. The clock signal typically originates
from the memory controller. This clock signal is connected to the bottom layer
of the 3D-stacked DRAM, and propagates toward the top layer. On the path to
this clock propagation, Cascaded-IO inserts one clock counter per layer, which,
when activated, can half the clock frequency.

Since these clock counters only perform simple divide-by-two, our clock path
can only generate frequency values split by powers of two. Therefore, the ideal
frequency for Layer~1, $3F$, cannot be generated. Instead, we must continue to
use a $4F$ frequency clock. Layer~2, however, can be driven at only $2F$, while
Layer~3 can be driven at frequency $F$. In general, the lower half of the $L$
layers will always run at frequency $F \times L$, while a quarter of them will
run at frequency $\frac{F \times L}{2}$, an eighth of them at frequency
$\frac{F \times L}{4}$, etc. The uppermost layer will always run at frequency
$F$. Figure~\ref{fig:cio_rclk} illustrates how the timings and data transfer
work with our reduced clock mechanism.\footnote{To avoid issues due to clock
skew, the data multiplexer for a layer is actually only responsible for
synchronizing its own data. Afterwards, when the multiplexer switches to
``receive'' data from the upper layer, it simply connects a bypass path from
the upper layer to the lower layer for the remaining cycles. This allows the
upper layer to directly send its data to the bottommost layer.
Figure~\ref{fig:cio_rclk} depicts the timing of this cut-through mechanism with
skew. For example, in Layer~1, the data switches from Layer~2's bit to
Layer~3's bit in the middle of the Layer~1 clock cycle, since the multiplexer
in Layer~1 is now directly connecting Layer~2 to Layer~0 without
synchronization.}

\subsubsection{Implementation Overhead} \label{sec:cio_overhead}

To implement Cascaded-IO-based IO interfaces, each DRAM chip needs to include
three components: {\em (i)} one clock counter, and {\em (ii)} a data
multiplexer for every TSV, and {\em (iii)} control logic to switch the
multiplexers between data sources (either connecting to the upper layer, or
driving the contents of its own cells) based on the clock. The counter and
multiplexer have very simple design, and only require a small number of
transistors, resulting in negligible area overhead. For example, in Layer~0
from Figure~\ref{fig:cio}, the multiplexer control consists of a two-bit
counter, which connects its IO drivers to the lower TSVs when the counter value
is {\tt 2'b00}. For all other values, the control will switch the multiplexer
to connect the input from the upper layer to the lower TSV.  Since only the
counter is necessary, the overhead and complexity of the control logic are also
negligible.

{\color{black}
\subsubsection{Power Consumption Overhead} \label{sec:cio_power}

Due to the different frequencies and data bandwidths at each layer, the bottom
layer in Cascaded-IO will indeed consume greater power. As the power network is
driven from the bottom, power delivery will be strongest in the lower layer of
3D-stacked DRAM, and significantly weaker in the upper
layers~\cite{shevgoor-micro13}. In these upper layers, Cascaded-IO provides a
new solution to reduce power consumption, which as a byproduct allows the
layers of Cascaded-IO to align well with the power network strength at each
layer. While the bottom layer operates at 4X frequency, the upper layers reduce
their clock frequencies to match their much lower bandwidth requirements (e.g.,
the topmost layer is clocked at 1X frequency). Therefore, while the physical
design of each layer remains homogeneous, we can minimize the increase in power
by introducing heterogeneity in the layers' operating frequencies. Such an
approach has much lower power consumption than if we were to run all of the
layers at 4X frequency (as is done in Dedicated-IO).

With its strong power delivery, the bottom layer can deliver similar per-pin
bandwidth as other high performance DRAMs (e.g., GDDR5, which has maximum $6
Gbps/pin$). It is important to note that a group of IO pins (usually $2 - 4$ IO
pins per group) has a dedicated power source (both {\tt Vddq} and {\tt Vssq}).
Therefore, it is reasonable to estimate the IO power capacity based on the IO
clock frequency (or per-pin bandwidth) rather than overall per-chip bandwidth.
Considering that, the bottom layer should operate as high as $6 Gbps/pin$
without power issues (as GDDR5 does), which is much higher than the highest IO
frequency ($800 Mbps/pin$), which we assume in our evaluations.
We discuss detailed power analysis in Section~\ref{sec:energy}.
}

\section{Rank Organizations with Multiple Layers} \label{sec:rank_org}

In DRAM, a rank is a collection of DRAM chips that operate in lockstep. A
rank-based organization allows pieces of a row to be split up across these
chips, since they logically appear to be a single structure. Existing
3D-stacked DRAM treats each layer as a single rank, as only one of these layers
can be accessed at a time. SMLA opens up new possibilities for the rank
organization, since multiple DRAM layers can now be accessed simultaneously.
In this section, we explore two such organizations: {\em Multi-Layer Ranks}
(MLR) and {\em Single-Layer Ranks} (SLR), which represent a key trade-off
between data transfer latency and rank-level parallelism.

\begin{figure}[hb]
	\vspace{-0.1in}
	\center
	\subfloat[{\bf SMLA} with {\bf Multi-Layer Ranks} (1 Rank, 2 Banks/Rank)] {
		\includegraphics[width=0.55\linewidth]{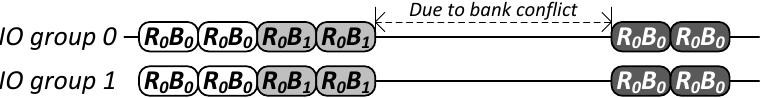}
		\label{fig:smla_mlr}
	}
	\vspace{-0.05in}

	\subfloat[{\bf Dedicated-IO} with {\bf Single-Layer Ranks} (2 Ranks, 2 Banks/Rank)] {
		\includegraphics[width=0.55\linewidth]{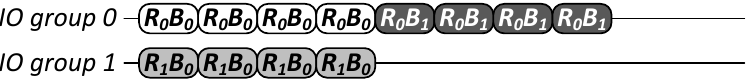}
		\label{fig:dio_slr}
	}
	\vspace{-0.1in}

	\subfloat[{\bf Cascaded-IO} with {\bf Single-Layer Ranks} (2 Ranks, 2 Banks/Rank)] {
		\includegraphics[width=0.55\linewidth]{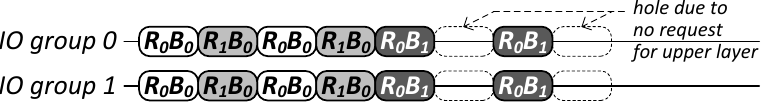}
		\label{fig:cio_slr}
	}
	\vspace{-0.05in}

	\subfloat{
		\includegraphics[width=0.55\linewidth]{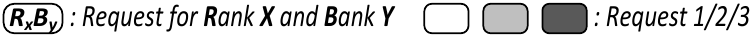}
	}

	\vspace{-0.05in}
	\caption{Request Service Timeline for Two Rank Designs}
	\vspace{-0.1in}
	\label{fig:timeline_rank}
\end{figure}

Figure~\ref{fig:timeline_rank} shows the timeline of serving requests using MLR
and SLR in a two-layer stacked DRAM, with either Dedicated-IO or Cascaded-IO.
To show the data flow of Dedicated-IO, the IO bus is divided into two groups
({\em IO Group 0} and {\em IO Group 1}) that are each connected to only one of
the two stack layers in Dedicated-IO (To compare Cascaded-IO and Dedicated-IO,
we keep using two IO groups in the timeline for Cascaded-IO, even though
Cascaded-IO does not need to divide IO bus into IO groups.) MLR merges both
layers into a single rank, while SLR provides two ranks. In both cases in our
example, a rank consists of two banks, to avoid rearchitecting the chips. (In
the case of MLR, this will result in banks with twice the capacity.) We
illustrate the timeline for serving three requests, each of which requires four
data transfers through an IO group (two data transfers through the entire IO).

{\bf SMLA with Multi-Layer Ranks.} MLR is similar to the current rank
organization of non-stacked DRAM, with multiple layers sharing a common command
interface. In our example MLR-based memory system with either Dedicated-IO or
Cascaded-IO, the data is transferred through both IO groups, requiring {\em
two} clock cycles to serve a request ({\em Request 1} in
Figure~\ref{fig:smla_mlr}). Note that while both Dedicated-IO and Cascaded-IO
have the same timeline when observing IO groups, the data served resides in two
different physical locations. In Dedicated-IO, the data for {\em IO Group 0}
comes from one layer (e.g., the bottom layer) and the {\em IO Group 1} data
comes from the other (top) layer. In Cascaded-IO, the first piece of data for
both IO groups come from the bottom layer, while the second piece comes from
the top layer. The third request in our sequence has an access to the same bank
and rank as the first request. Therefore, there will be a delay before serving
the third request due to a bank conflict~\cite{salp}.

{\bf SMLA with Single-Layer Ranks.} In a Dedicated-IO-based memory system with
SLR (Figure~\ref{fig:dio_slr}), each layer is a rank, and has its own IO group.
Therefore, a request to a rank will only transfer data over its assigned group,
requiring four clock cycles to serve a request. On the other hand, Cascaded-IO
partitions its data transfer cycles in a round robin fashion, with each layer
having an assigned time slot (one every $L$ cycles). In
Figure~\ref{fig:cio_slr}, {\em Rank 0} corresponds to the bottom layer, and
{\em Rank 1} to the top layer. We assign the lower layer to the first of the
two time slots. Therefore, the odd data bursts correspond to {\em Rank 0}
requests, while the even data bursts correspond to {\em Rank 1} requests. In
this way, the first and third data bursts represent the first complete request
to {\em Rank 0}. When our third request is serviced, since the Cascaded-IO
slot assignments are fixed, there will be a hole in between its two data
bursts, since {\em Rank 1} will not be serving anything.

{\bf Trade-Off: MLR vs.\ SLR.} For our two-layer memory system, MLR can fully
service an individual request within two cycles, through the use of both IO
groups. In contrast, SLR takes up to four cycles to service each request, but
can deliver \emph{two} requests in that time, since it can overlap latencies
for accesses to different ranks. MLR therefore better supports latency-bound
applications that do not issue large numbers of memory requests. SLR, on the
other hand, is better tuned for memory-level parallelism, as it exposes a
greater number of ranks. As we can see in Figure~\ref{fig:timeline_rank}, the
third request experiences fewer delays in SLR because the larger number of
available ranks reduces the probability of conflicts.

While we have considered two extremes of rank organization (one layer per rank
vs. using all layers in a rank), it is possible to pick a design point in the
middle (e.g., two ranks in a four-layer memory, with two layers per rank). We
leave the evaluation of such organizations as future work.

\section{Energy Consumption Analysis} \label{sec:energy}

In order to deliver greater memory bandwidth, SMLA increases the memory channel
frequency, which can change energy consumption significantly. To analyze the
energy efficiency of our mechanisms, we: {\em (i)} observe the trend of DRAM
energy consumption at different operating frequencies, {\em (ii)} determine
which components of energy consumption are coupled with clock frequency, and
{\em (iii)} estimate the overall energy consumption of our proposed mechanisms.

\begin{figure}[ht]
	\centering
	\includegraphics[width=0.65\linewidth]{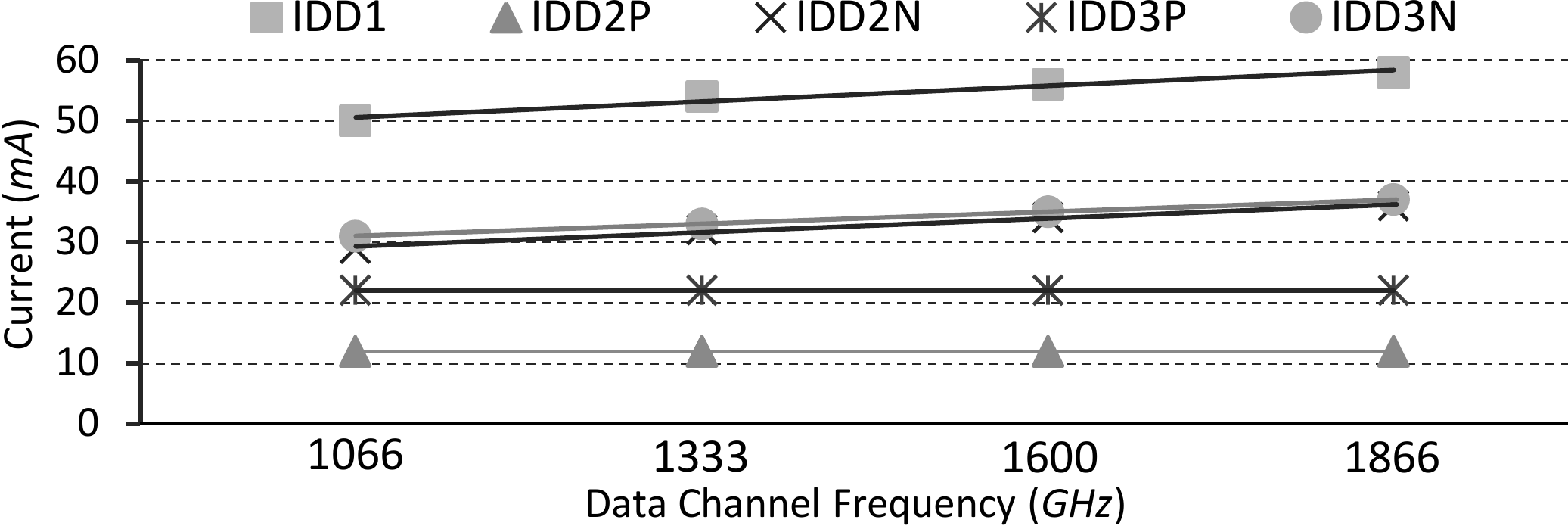}
	\caption{DDR3 Current Trend over Channel Frequency}
	\label{fig:trend_current}
\end{figure}

Figure~\ref{fig:trend_current} plots the current specifications of DDR3 at four
different IO frequencies ($1066$, $1333$, $1600$, and $1866MHz$, based on
manufacturer specifications~\cite{micron_spec}). Each current represents a
specific property of DRAM --- either periodically-performed operations ({\tt
IDD1}) or current due to staying in a particular state ({\tt IDD2N/P} and {\tt
IDD3N/P}).

In power-down mode ({\tt IDD2P} and {\tt IDD3P}), DRAM exploits current
reduction techniques that stop the internal clock, resulting in no variation at
different frequencies. Standby currents at different frequencies ({\tt IDD2N}
and {\tt IDD3N}) show large variation (about a $20-30\%$ energy reduction when
cutting the frequency in half). Therefore, DRAM consumes a significant fraction
of overall energy to maintain the clock propagation network. When accessing
DRAM cells, the absolute amount of the current reduction from halving the
frequency is similar to the reduction in standby mode. However, due to the high
energy consumption when accessing cells (either activate/precharge or
read/write), the amount of reduced energy is relatively small (about $10-15\%$)
compared to the reduction in standby modes. We draw two observations from these
results. First, increasing the clock frequency leads to greater current flow
(larger energy consumption), with the relationship between energy consumption
and clock frequency mostly linear. Second, the slope of the current-frequency
trend is similar in normal mode for all frequencies.

Using this observation along with DRAM energy estimation
tools~\cite{rambus-power10,vogelsang-micro10,micron-power}, we extract two
current sources: {\em (i)} current coupled with the clock, and {\em (ii)}
current decoupled from the clock. We then estimate the current consumptions at
different frequencies. Table~\ref{tbl:current} presents energy consumption for
three different operating conditions (power-down, standby, and access) at four
different operating frequencies ($200$, $400$, $800$, and $1600MHz$), estimated
from the measured energy consumption of 3D-stacked
DRAM~\cite{chandrasekar_date13}. We see that a large fraction of DRAM energy
consumption in standby modes is coupled with the operating clock frequency.

\begin{table}[hb]
	\vspace{-0.05in}
	\centering
{\small
	\begin{tabular}{lcccc}
	\toprule
		 Clock Frequency ($MHz$)									& 200  & 400  & 800  & 1600 \\ \midrule
		 Power-Down Current ($mA$)								& 0.24 & 0.24 & 0.24 & 0.24  \\ 
		 Precharge-Standby Current ($mA$)					& 4.24 & 5.39 & 6.54 & 8.84 \\ 
		 Active-Standby Current ($mA$)						& 7.33 & 8.50 & 9.67 & 12.0 \\ 
		 Active-Precharge without Standby ($nJ$)	& 1.36 & 1.37 & 1.38 & 1.41 \\ 
		 Read without Standyby ($nJ$)	 						& 1.93 & 1.93 & 1.93 & 1.93 \\ 
		 Write wirhout Standyby ($nJ$) 						& 1.33 & 1.33 & 1.33 & 1.33 \\ \bottomrule
	\end{tabular}
}
	\vspace{-0.05in}
	\caption{Energy Consumption Estimation}
	\vspace{-0.1in}
	\label{tbl:current}
\end{table}

\section{Methodology} \label{sec:method}

We use a cycle-accurate in-house simulator that models 3D-stacked memory and
whose front-end is based on Pin~\cite{pin,pinpoints}.
{\color {black}
We use existing 3D-stacked DRAM, WideIO, as our baseline. More recent
3D-stacked DRAM proposals (WideIO2, HMC, HBM) can enable more bandwidth at the
expense of area and power consumption (as shown in
Section~\ref{sec:bandwidth}). As we discussed, the key advantage to our
mechanisms (Dedicated-IO and Cascaded-IO) is our ability to deliver competitive
bandwidth to these high-bandwidth proposals at a much lower global sense
amplifier cost. It is also important to note, however, that simply performing
bandwidth scaling (as WideIO2/HMC/HBM have done) will eventually be constrained
by the area and power overhead of these extra global sense amplifiers. We show
that our mechanisms overcome this challenge with a large reduction in
sense amplifier count, and can still be used on top of these more
recently-proposed DRAM architectures to further increase off-chip bandwidth.
}

We evaluate our mechanisms for 3D-stacked DRAM with two/four/eight layers, with
a corresponding increase in IO frequency of 2/4/8 times, respectively, over
conventional 3D-stacked DRAM. We evaluate two rank organizations (Single-Layer
Rank and Multi-Layer Rank). Table~\ref{tbl:memory} summarizes the parameters
for the baseline 3D-stacked memory system and our proposed memory systems,
which consist of four stacked layers and operate at 4X of the baseline system's
IO frequency.

\begin{table}[hb]
\vspace{-0.05in}
	\centering
{\small
	\begin{tabular}{cccccc}
	\toprule
		 IO interface 								& Baseline  & \multicolumn{2}{c}{Dedicated-IO} &
	 																\multicolumn{2}{c}{Cascaded-IO}					\\ 
		 Rank Organization 						& SLR & MLR & SLR & MLR & SLR						\\ \midrule
		 Number of Ranks 							& 4 	& 1 	& 4 	& 1 	& 4							\\ 
		 Clock Freq.~({\em MHz})			& 200	& 800	& 800 & 800 & 800						\\ 
		 Bandwidth~({\em GBps})				& 3.2	& 12.8& 12.8& 12.8& 12.8					\\ 
		 Data Trans. Time~({\em ns})	& 20.0& 5.0 & 20.0& 5.0	& 18.1$\dagger$	\\ 
		 Sim. Multi Layer Acc. 				& 1		& 4 	& 4		& 4		& 4							\\ \bottomrule \\
	\end{tabular}
	\vspace{-0.1in}
	\\ $\star$ Global Param.: $4$ $Layers$, $2$ $Banks/Rank$, $128$ $IOs/Chan.$, $64$ $Byte/Req.$\\
	$\dagger$ Avg. Data Transfer Time of Layers ($Bottom:16.25/17.5/18.75/Top:20$)
}

\vspace{-0.05in}
\caption{3D-Stacked DRAM Configurations Evaluated} \label{tbl:memory}
\vspace{-0.05in}
\end{table}

Table~\ref{tbl:system} shows the evaluated system configuration. We use a one
channel memory system for our single-core evaluations, and a four channel
system for multi-core evaluations (identical to the configuration used in
Wide-IO~\cite{jedec-wideio}).

\begin{table}[ht]
	\vspace{-0.05in}
	\centering
{\small
 	\begin{tabular}{lp{5.5cm}}
		\toprule
		Component & Parameters \\
		\midrule
		\multirow{2}{*}{Processor} 				& 1 -- 16 cores, 3.2GHz, 3-wide issue, \\
																			& 8 MSHRs, 128-entry instruction window \\
		\multirow{2}{*}{Last-Level Cache} & 64B cache-line, 16-way associative, \\
         															& 512KB private cache-slice per core \\
		\multirow{1}{*}{Memory} 					& 64/64-entry read/write queues/controller,\\
		\multirow{1}{*}{Controller}				& FR-FCFS scheduler~\cite{frfcfs} \\
		\multirow{1}{*}{Memory System} 		& 2 -- 8 layer stacked DRAM, 1 -- 4 channel\\
		\bottomrule
	\end{tabular}
}
	\vspace{-0.05in}
	\caption{Configuration of Simulated System} \label{tbl:system}
	\vspace{-0.05in}
\end{table}

We use 31 applications from the SPEC CPU2006, TPC~\cite{tpc}, and
STREAM~\cite{stream} applications suites. For single-core studies, we report
results that are averaged across all applications. For our multi-core
evaluation, we generate 16 multi-programmed workloads for each case (4--16
cores / 2--8 stacked layers) by randomly selecting from our workload pool.

We execute all applications for 100 million instructions, similar to many prior
works~\cite{vwq,stfm-micro07,microarch-mlp,salp} that also studied memory
systems. For multi-core evaluations, we ensure that even the slowest core
executes 100 million instructions, while other cores continue to exert pressure
on the memory subsystem. To measure performance, we use instruction throughput
(IPC) for single-core systems and {\it weighted speedup}~\cite{ws} for
multi-core systems.

\section{Evaluation} \label{sec:result}

In our experiments for both single-core and multi-core systems, we compare
three different 3D-stacked DRAM designs: {\em (i)} the baseline conventional
3D-stacked DRAM, {\em (ii)} \emph{Dedicated-IO}, {\em (iii)} \emph{Cascaded-IO}
(the latter two are SMLA-based designs) both in terms of performance and energy
efficiency.

\subsection{Single-Core Results} \label{sec:result_single}

\begin{figure}[ht]
	\center
	\vspace{-0.1in}
	\subfloat[Multi-Layer Rank (1 rank per channel)] {
		\begin{tabular}{c}
			\includegraphics[width=0.45\linewidth]{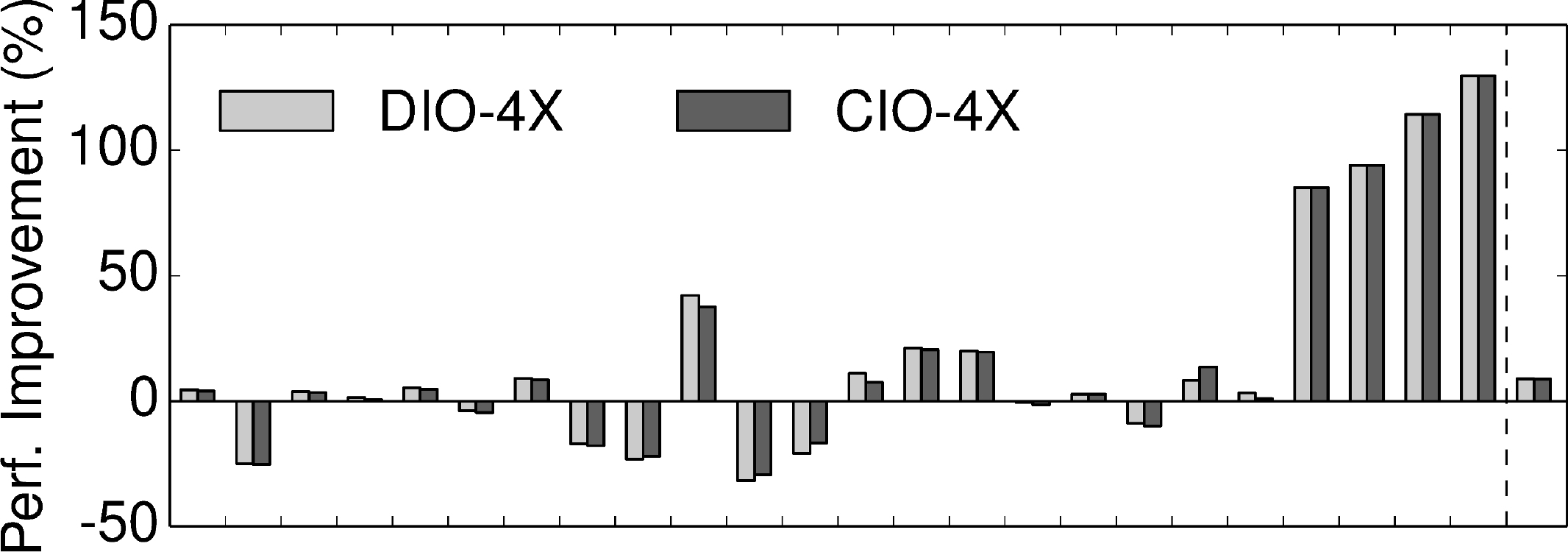} \\
			\includegraphics[width=0.45\linewidth]{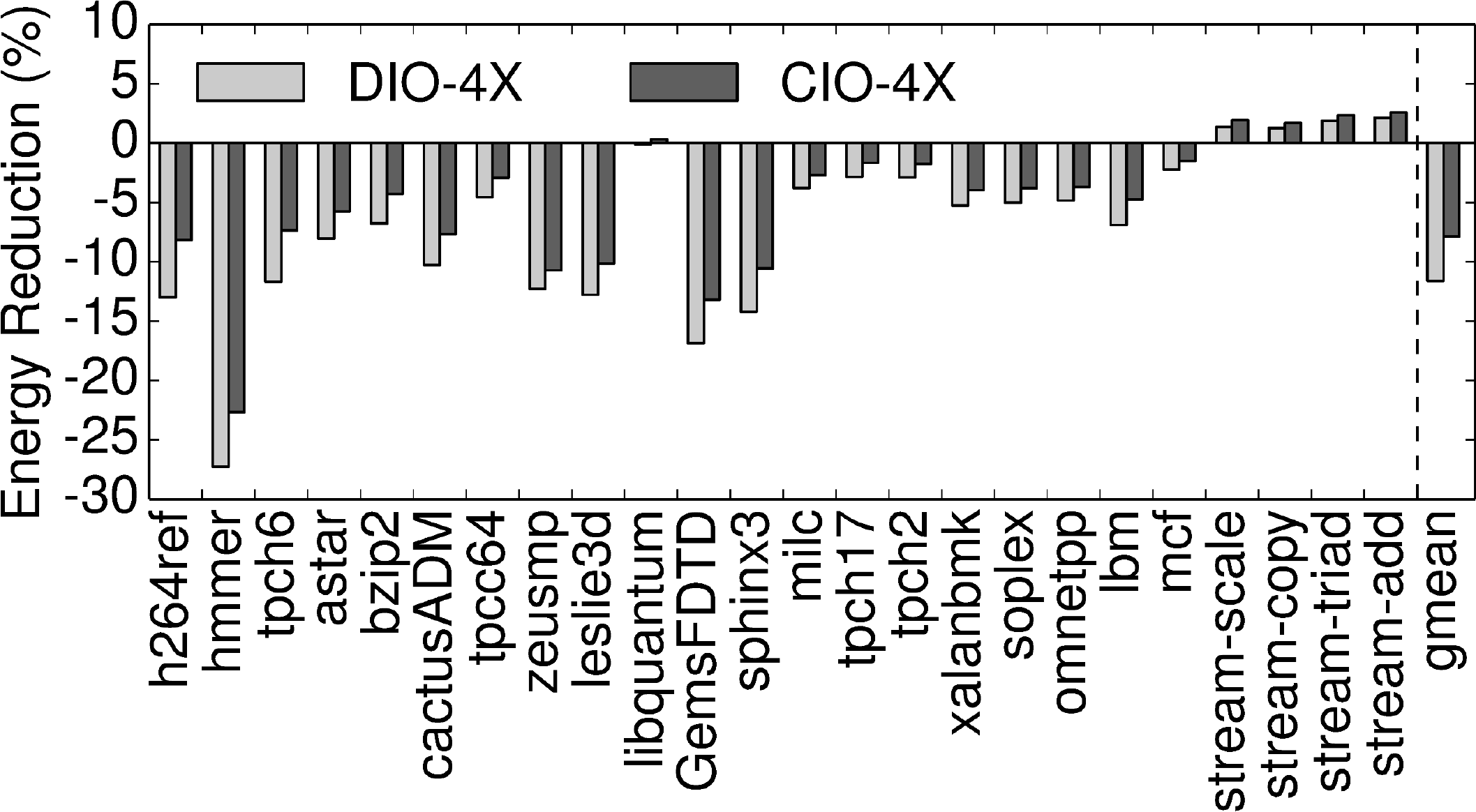}
		\end{tabular}
		\label{fig:result_single_mlr}
	}
	\subfloat[Single-Layer Rank (4 ranks per channel)] {
		\begin{tabular}{c}
			\includegraphics[width=0.45\linewidth]{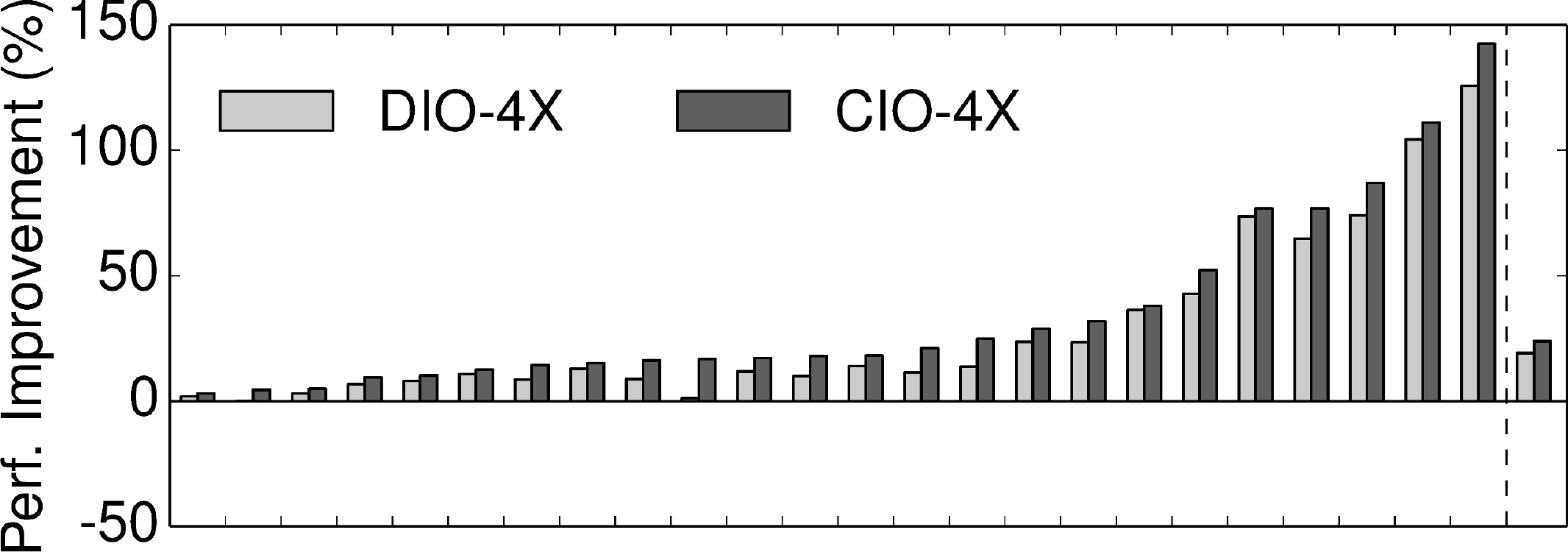} \\
			\includegraphics[width=0.45\linewidth]{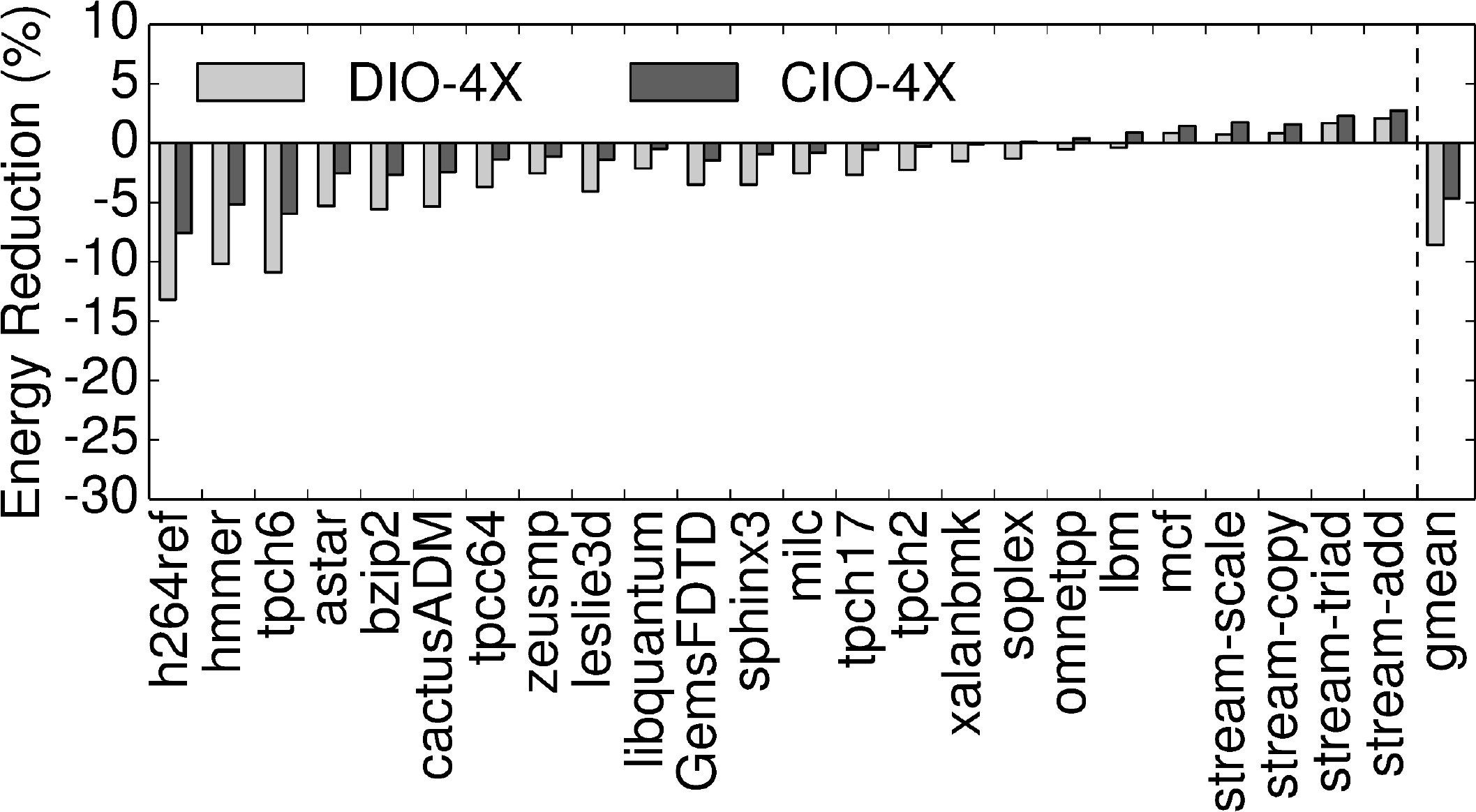}
		\end{tabular}
		\label{fig:result_single_slr}
	}
	\vspace{-0.05in}
	\caption{Performance and Energy Efficiency Evaluations on a Single-Core System}
	\label{fig:result_single}
	\vspace{-0.1in}
\end{figure}

Figure~\ref{fig:result_single} compares SMLA-based designs (Dedicated-IO and
Cascaded-IO) to the baseline with conventional 3D-stacked DRAM.\footnote{We
present 24 applications whose $MPKI$ (last-level cache
Misses-Per-Kilo-Instructions) are larger than 1. The other benchmarks ($MPKI <
1$) show performance changes of less than 1\%. However, the provided average (geometric
mean) results for both performance and energy include all 31 workloads.} All
three 3D-stacked DRAM designs have four stacked layers
(Table~\ref{tbl:memory}). By accessing those four layers simultaneously,
SMLA-based mechanisms provide four times more bandwidth compared to the
baseline. Each mechanism was evaluated with the two rank organizations:
Single-Layer Rank (SLR, 4 ranks per channel in
Figure~\ref{fig:result_single_slr}), and Multi-Layer Rank (MLR, 1 rank per
channel in Figure~\ref{fig:result_single_mlr}). For each application, the
figures show two metrics: {\em (i)} performance improvement of SMLA-based
mechanisms compared to the baseline, and {\em (ii)} the power reduction of
SMLA-based mechanisms relative to the baseline. We draw three conclusions from
the figure.

{\bf Performance with Rank Organizations.} First, {\em SMLA-based mechanisms
provides significant performance improvement on average in both rank
organizations. However, individual application shows different performance
benefit based on the rank organizations.}

In SLR (Figure~\ref{fig:result_single_slr}), for most applications, both of our
mechanisms improve performance significantly compared to the baseline. On
average, Dedicated-IO and Cascaded-IO improve performance by 19.2\% and 23.9\%,
respectively, compared to the baseline. Intuitively, performance benefits are
higher for applications with higher memory intensity (MPKI of the seven
left-most applications is less than 8, while the average MPKI of the seven
right-most applications is more than 35.).

In MLR (Figure~\ref{fig:result_single_mlr}), while SMLA-based mechanisms
improve performance for most applications compared to the baseline (by 8.8\% on
average for both mechanisms), there are a few cases performance degrades. Note
that even though both Dedicated-IO and Cascaded-IO enable much higher bandwidth
and lower latency, these mechanisms with MLR have only one rank per channel.
This reduces the number of opportunities for rank-level parallelism. As a
result, several applications (e.g., {\tt hmmer, zeusmp, leslie3d, GemsFDTD,
sphinx3, omnetpp}) will benefit more from having more rank-level parallelism
(in the baseline) than from having higher bandwidth and lower latency (in MLR).

For most applications, SMLA-based mechanisms with SLR provide better
performance improvement compared to the corresponding benefits with MLR. There
are, however, a few noticeable exceptions (e.g., {\tt libquantum, h264ref}) --
applications where Dedicated-IO and Cascaded-IO show better performance
improvement with MLR. These applications are more latency sensitive, and hence
they benefit more from low-latency in MLR than from rank-level parallelism in
SLR.

{\bf Performance with SMLA Designs.} Second, {\em Cascaded-IO usually provides
better performance improvement than Dedicated-IO in SLR.} This is because
Cascaded-IO has lower average latency than Dedicated-IO in SLR. However, in
MLR, both mechanisms have same latency (shown in Table~\ref{tbl:memory}).
Furthermore, upper layers in Cascaded-IO operate at lower frequency, leading to
reduction in the command and address bandwidth for the upper layers. As a
result, in MLR, Dedicated-IO shows a slightly better performance than
Cascaded-IO.

{\bf Energy Consumption.} Third, {\em Cascaded-IO shows better energy
efficiency compared to Dedicated-IO due to reduction in the frequency of the
third and fourth layers.} However, due to the increase in the overall
frequency, on average (and especially for lower layers), both Dedicated-IO and
Cascaded-IO consume more energy compared to the baseline (8.6\% and 4.6\% more
with SLR and 11.6\% and 7.8\% with MLR, respectively).

\subsection{Multi-Core Results} \label{sec:result_multi}

\begin{figure}[ht]
	\hspace{0.2in}
	\vspace{-0.1in}
	\center
	\subfloat[Performance Improvement] {
		\includegraphics[width=0.3\linewidth]{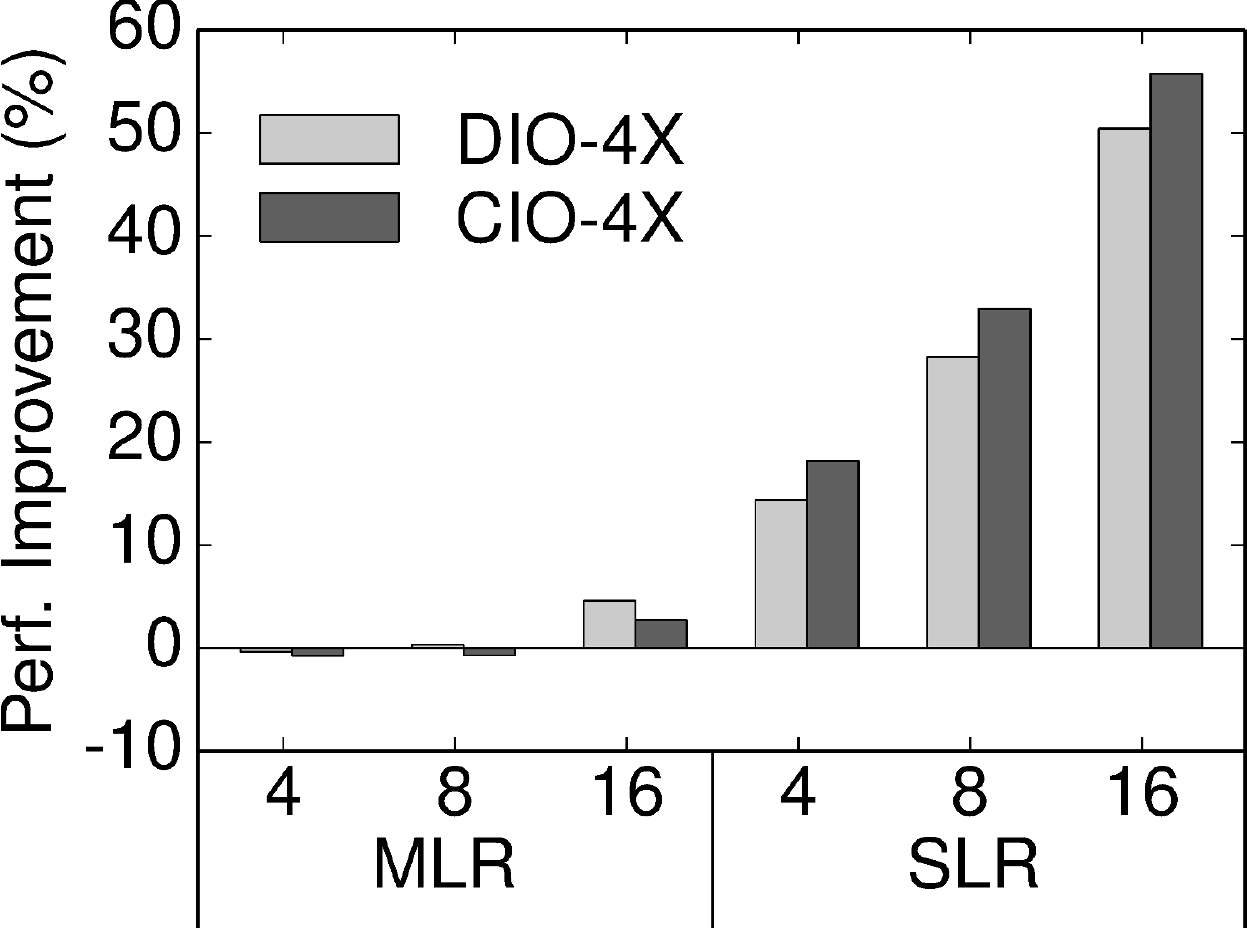}
		\label{fig:result_ipc_multi}
	} \hspace{0.2in}
	\subfloat[Energy Reduction] {
		\includegraphics[width=0.3\linewidth]{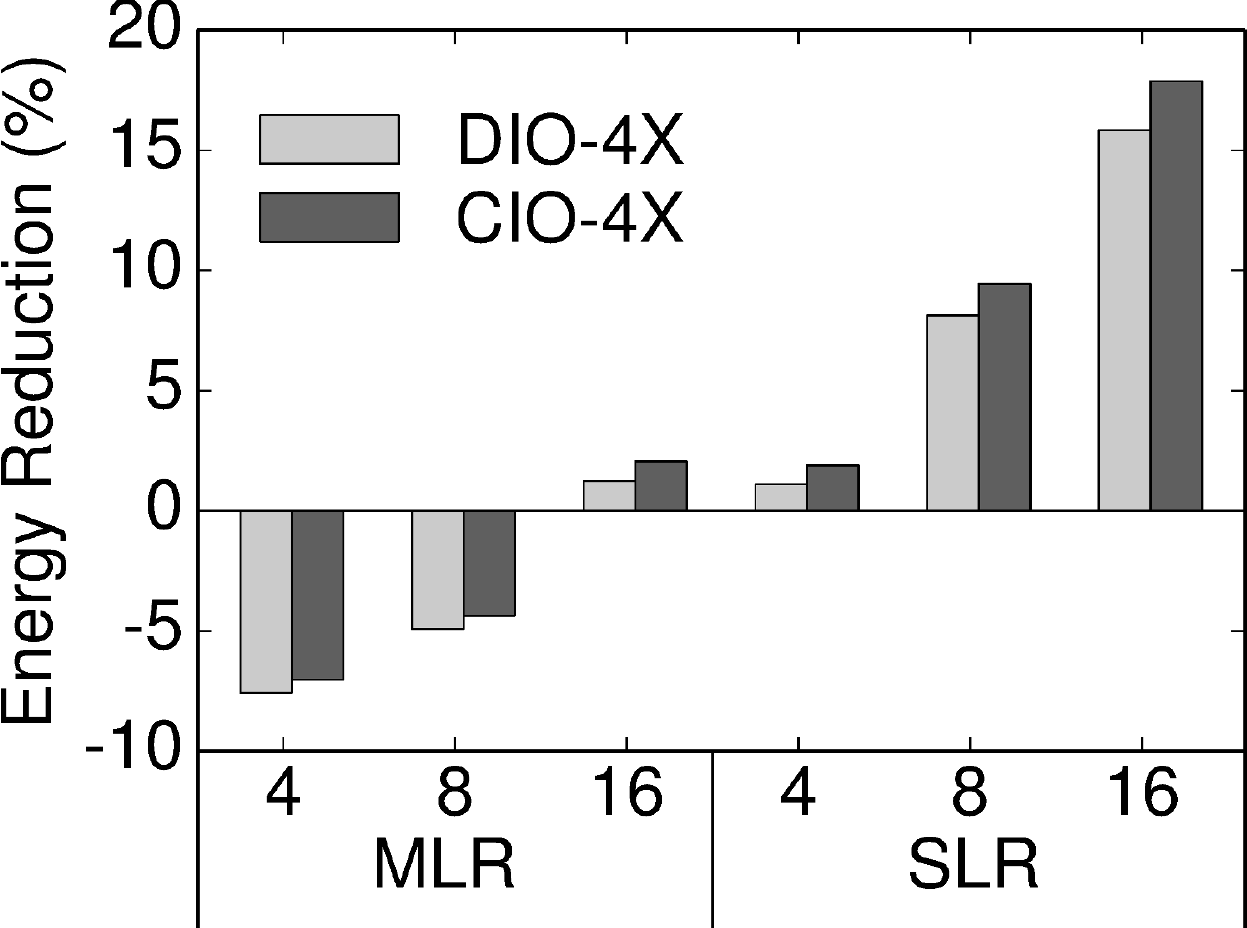}
		\label{fig:result_pwr_multi}
	}
	\vspace{-0.05in}
	\caption{Multi-Core Evaluation (4/8/16 Cores, 4 Layers)}
	\label{fig:result_multi}
	\vspace{-0.1in}
\end{figure}

Figure~\ref{fig:result_ipc_multi} shows the system performance improvement for
our mechanisms in multi-programmed workloads. Similar to single-core results
with SLR, both Dedicated-IO and Cascaded-IO provide significant performance
improvement in all system configurations. On average, Dedicated-IO obtains a
14.4\%/28.3\%/50.4\% weighted speedup for 4/8/16-core systems, respectively,
while Cascaded-IO provides a further performance bump of 18.2\%/32.9\%/55.8\%.
However, due to reduction in rank-level parallelism, the performance benefits
of our mechanisms using MLR are lower than with SLR.

{\color{black}
Our mechanisms increase power consumption (energy/time) due to increased IO
frequency and enabling more IO bandwidth. However, our mechanisms can avoid
overall memory system energy increase, mainly due to reduced execution time in
our applications.
}
We see that unlike in the single-core case, our mechanisms with SLR can
actually deliver significant energy reductions as the number of cores (and
therefore the memory contention) increases. In such cases, where there is a
higher bandwidth demand, SMLA alleviates much of this contention, resulting in
shorter overall execution times. Figure~\ref{fig:result_pwr_multi} shows these
savings with SLR, which on average are 1.9\%/9.4\%/17.9\% for 4/8/16 cores,
respectively. Unlike SLR, MLR is geared towards latency and not parallelism,
and as such is unable to ease the contention pressure as well, resulting in
increased energy consumption.

\subsection{Sensitivity to the Number of Stacked Layers} \label{sec:layer}

\begin{figure}[ht]
	\hspace{0.2in}
	\vspace{-0.2in}
	\center
	\subfloat[Performance Improvement] {
		\includegraphics[width=0.3\linewidth]{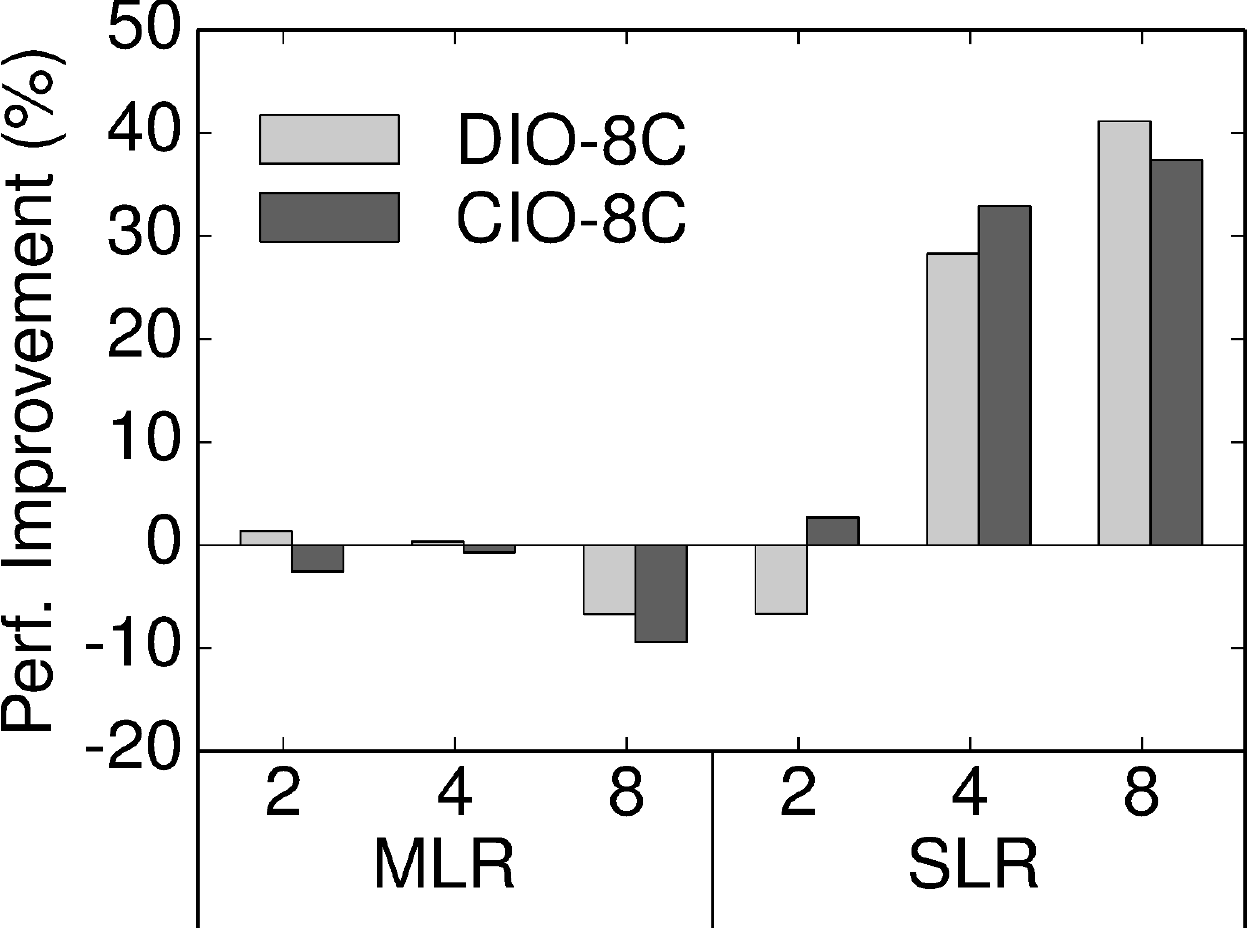}
		\label{fig:result_ipc_layer}
	}\hspace{0.2in}
	\subfloat[Energy Reduction] {
		\includegraphics[width=0.3\linewidth]{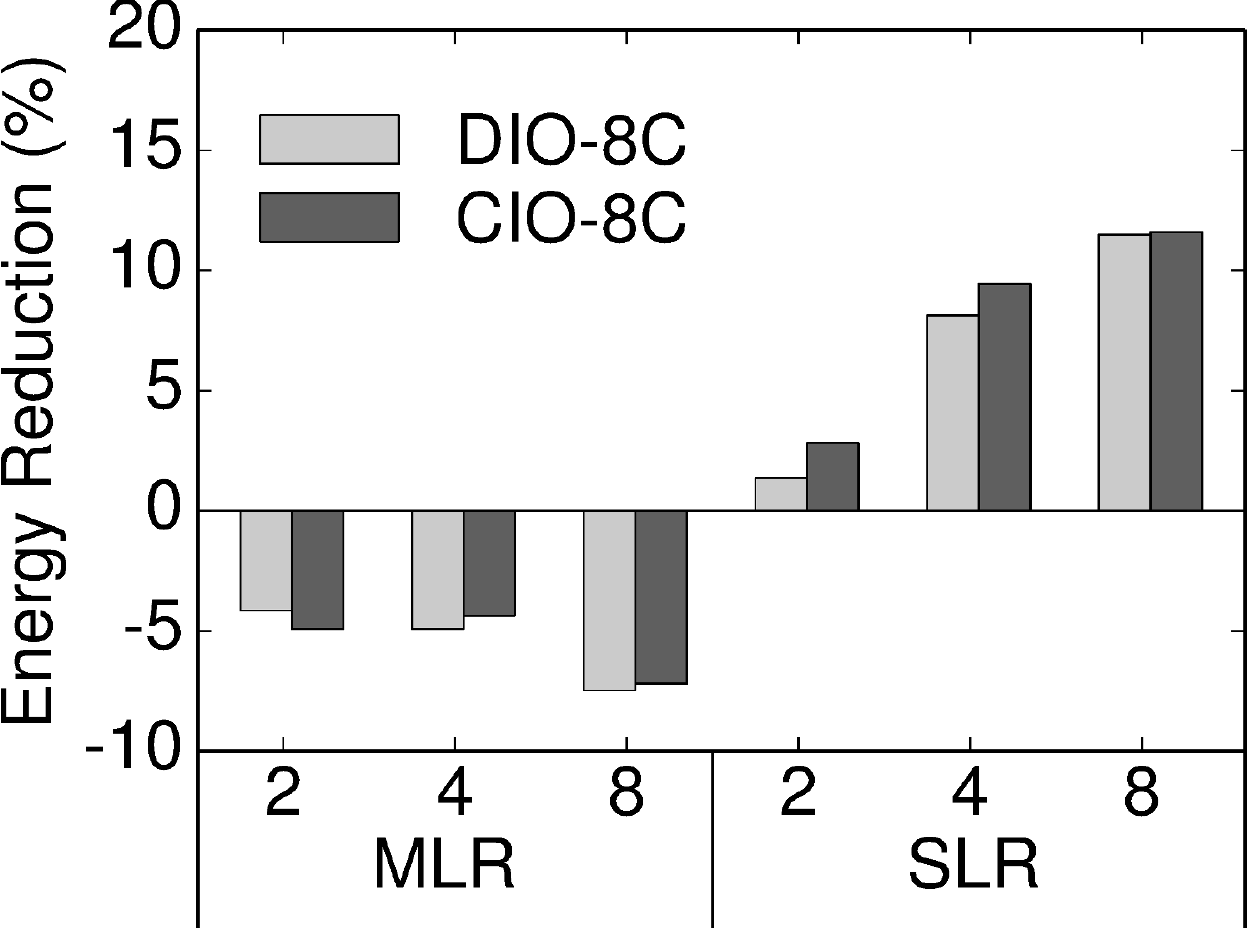}
		\label{fig:result_pwr_layer}
	}
	\vspace{-0.05in}
	\caption{Sensitivity for Layer-Count (8 Cores, 2/4/8 Layers)}
	\label{fig:result_layer}
	\vspace{-0.1in}
\end{figure}

The maximum bandwidth improvement in our mechanisms depends on the number of
stacked-layers whose global bitline interfaces can operate simultaneously. For
example, in a two layer-stacked memory system, our mechanism can enable twice
the bandwidth of the conventional 3D-stacked DRAM (4/8 times the bandwidth for
4/8 layer-stacked DRAM, respectively). Figure~\ref{fig:result_layer} shows the
system performance improvement and energy reduction of our mechanisms in 2 -- 8
layer-stacked DRAM. We use eight-core multi-programmed workloads and our
evaluated memory system has four channels.

Two observations are in order. First, as expected, our mechanisms with SLR
provide better performance and energy efficiency that grows with the number of
stacked layers in DRAM. At the same time, we observe less performance benefits
in eight layer-stacked DRAM with MLR, mainly due to reduction in the number of
ranks (one rank in MLR-based system vs. eight ranks in the baseline system).
Second, a 3D-stacked DRAM with more layers, (especially, eight layer-stacked
DRAM), has better performance with Dedicated-IO than with Cascaded-IO due to
reduction in frequency of the higher layers in Cascaded-IO. This, in turn,
leads to reduction in command bandwidth of upper layers.

\subsection{Energy Consumption with Varying Memory Intensity}
\label{sec:result_energy}

\begin{figure}[ht]
	\center
	\vspace{-0.1in}
	\subfloat[Normalized Energy Consumption] {
		\includegraphics[width=0.6\linewidth]{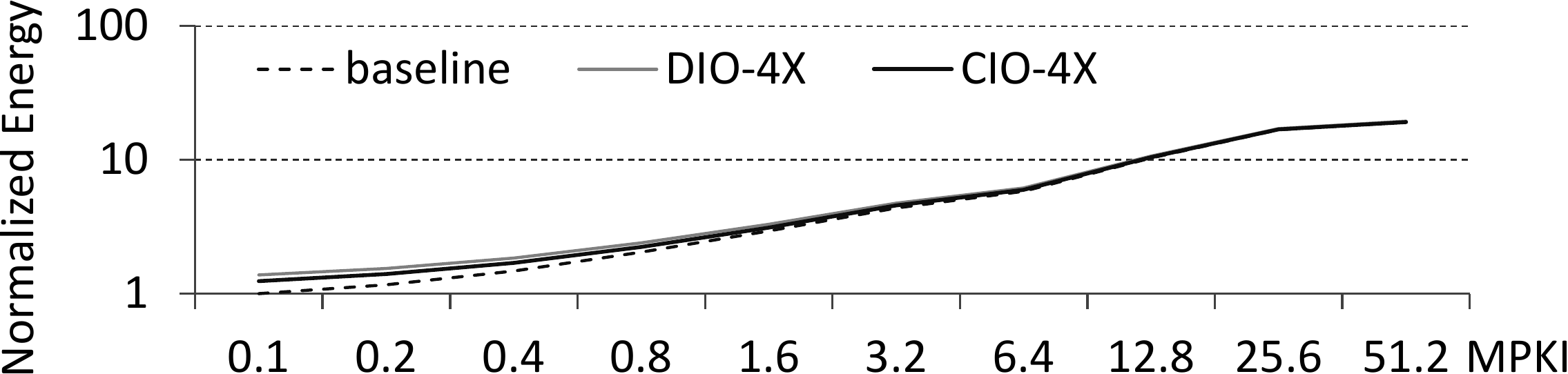}
		\label{fig:energy_normalized}
	}

	\subfloat[Energy Increase] {
		\centering
		\includegraphics[width=0.6\linewidth]{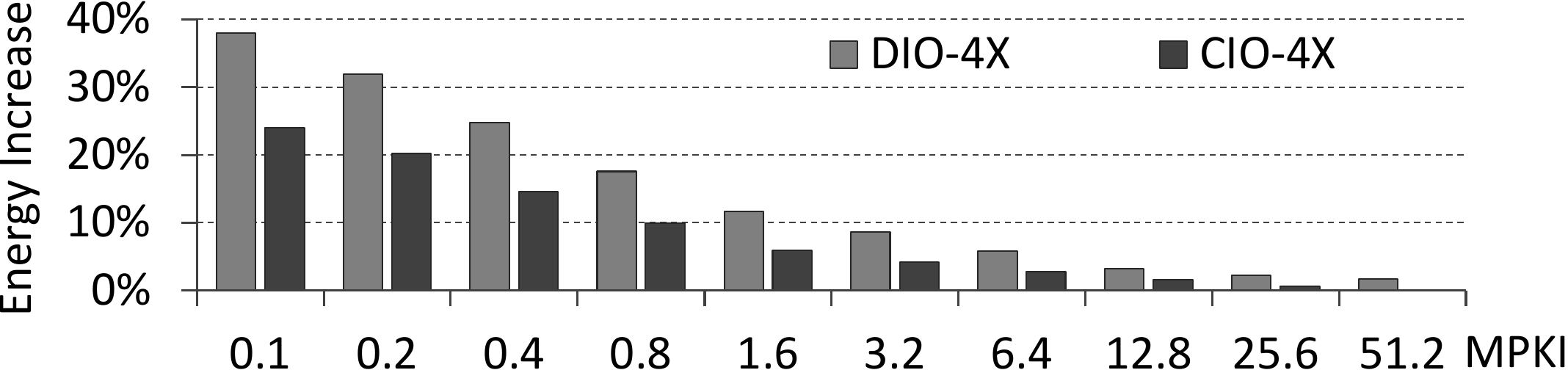}
		\label{fig:energy_relative}
}
	\vspace{-0.05in}
	\caption{Memory Intensity vs. Energy Consumption}
	\label{fig:result_energy}
	\vspace{-0.1in}
\end{figure}

Figure~\ref{fig:result_energy} shows the energy consumption of SMLA-based
mechanisms, compared to the baseline, with varying memory intensity (MPKI) of
micro-benchmarks in two different ways. First, we observe the absolute energy
consumption, which is normalized to the baseline with 0.1 MPKI
(Figure~\ref{fig:energy_normalized}). As expected, the energy consumption of
all three mechanisms is growing with increase in memory intensity (21X more
energy consumed at 51.2 MPKI than at 0.1 MPKI). Second, we analyze the relative
energy increase when executing the same applications
(Figure~\ref{fig:energy_relative}). We observe that the amount of energy
increase (relative to the baseline) significantly reduces at higher MPKIs. This
is because DRAM consumes more energy to serve more memory requests, and the
amount of energy consumed by these operations is not dependent on IO frequency.
In other words, the absolute value of additionally consumed energy is fairly
constant across all operating conditions, and relatively small compared to the
amount of energy to serve memory requests.

In all the cases, Cascaded-IO provides better energy efficiency than
Dedicated-IO (about 30\% lower energy overhead due to lower frequency of
certain layers). Based on this analysis, we conclude that {\em (i)} SMLA-based
deigns can increase energy consumption, but this overhead is relatively small
to the overall DRAM energy when running memory intensive applications, and it
is also small compared to the overall system energy when applications are not
memory intensive, and {\em (ii)} Cascaded-IO provides better energy efficiency
than Dedicated-IO.

\section{Related Work} \label{related}

To our knowledge, this is the first work that {\em (i)} enables higher
bandwidth at low cost by leveraging the existing global bitline interfaces in
multiple layers of 3D-stacked DRAM, and {\em (ii)} provides higher bandwidth
with low energy consumption by optimizing IO frequency individually for each
layer. In this section, we discuss the prior works that aim to improve memory
system bandwidth.

{\bf Bank-Group.} LPDDR3~\cite{jedec-lpddr3} and DDR4~\cite{jedec-ddr4}
categorize banks into multiple groups (bank-group), where each group owns a set
of global sense amplifiers. LPDDR3 and DDR4 increase DRAM internal bandwidth by
simultaneously accessing each bank-group, where our design provides higher
bandwidth in 3D-stacked DRAM by aggregating the DRAM internal bandwidth in
multiple layers. Potentially, these two approaches are orthogonal and can be
applied together to further increase the bandwidth in 3D-stacked DRAM.

{\bf High Bandwidth Memory.} HBM~\cite{jedec-hbm,lee-isscc14} enables high
bandwidth ($128GBps$) 3D-stacked DRAMs by {\em (i)} increasing DRAM internal
bandwidth by adding extra global bitlines per chip (2X compared to Wide-IO),
{\em (ii)} allowing simultaneous accesses to different bank-groups having their
own set of global sense amplifiers (same as~\cite{jedec-lpddr3,jedec-ddr4}),
and {\em (iii)} aggregating the bandwidth of each layer by assigning exclusive
channels to each layer. On the other hand, we provide higher bandwidth without
adding any extra bitline interface, by aggregating data available at different
layers and transferring them at a higher frequency. HBM can achieve performance
and energy efficiency similar to Dedicated-IO, but lower than Cascaded-IO,
while integrating 2X global sense amplifiers than our mechanisms (as shown in
Section~\ref{sec:bandwidth}).

{\bf 3D-Stacked DRAM Studies.} Prior works on 3D-stacked DRAM focused on
utilizing the higher capacity of it as cache, and main
memory~\cite{black-micro08,loh-micro09,woo-hpca10}. A prior study focused on
the architectural trade-offs in rank and channel
organizations~\cite{loh-isca08}. However, none of these works focus on solving
the limited internal bandwidth of DRAM in presence of wider IO.

{\bf Reconfiguring DIMM Organization.} Mini-rank~\cite{zheng-micro08} enables
independent accesses to sub-divided ranks, similar to our mechanisms. However,
our mechanisms enable more bandwidth by mitigating internal bandwidth
bottleneck of 3D-stacked DRAM, which is not possible in Mini-rank. By
integrating a buffer on DIMM, Decoupled-DIMM~\cite{zheng-isca09} enables higher
frequency memory channel by decoupling memory channel from DRAM chips, similar
to our mechanisms. However, our mechanisms can be implemented with small
changes in 3D-stacked DRAM, while Decoupled-DIMM requires the expensive high
performance buffer.

{\color{black} {\bf Multi-Layer Rank Organization.} Loh~\cite{loh-isca08}
introuduced a 3D-stacked DRAM that form a rank over multiple layers. That work
assumes that all upper layers only have DRAM cells and that the bottom layer
contains special logic to control all of the upper layers, which is not present
in existing 3D-stacked DRAMs. However, Our mechanisms enable MLR within
existing 3D-stacked DRAM architectures (where all layers have control logic and
communicate only across IO) by enabling simultaneous multi-layer access,
through minimal, uniform additions to the control logic.  }

\section{Conclusion}

In this work, we introduce Simultaneous Multi Layer Access (SMLA), a new IO
organization for 3D-stacked DRAM that enables greater off-chip memory bandwidth
at low cost. We identify that the major bandwidth bottleneck of existing
3D-stacked DRAM is the global bitline interface. We offer a cheaper
alternative to adding more global bitlines, by instead exploiting the
otherwise-idle internal bandwidth of multiple layers to drive a faster IO
interface.

We study two IO implementations for SMLA, including Cascaded-IO, which
time-multiplexes the IO bus across all of the DRAM layers and exploits its
dataflow to reduce the energy impact of a faster-clocked memory. We evaluate
SMLA over a wide array of applications, and show that our proposed mechanisms
significantly improve performance and reduce energy consumption (on average
55\%/18\%, respectively, for 16-core multi-programmed workloads) over
conventional 3D-stacked DRAMs. We conclude that SMLA provides a high
performance and energy efficient IO interface for building modern (and future)
3D-stacked memory systems at low cost.

\section*{Acknowledgments}

We thank the SAFARI group members for the feedback and stimulating research
environment they provide. We thank Uksong Kang, Jung-Bae Lee, and Joo Sun Choi
from Samsung for their helpful comments. We acknowledge the support of our
industrial partners: Facebook, Google, IBM, Intel, Microsoft, Qualcomm, VMware,
and Samsung. This research was partially supported by NSF (grants 0953246,
1212962, 1065112, 1320531), Semiconductor Research Corporation, and the Intel
Science and Technology Center for Cloud Computing. Donghyuk Lee is supported in
part by a Ph.D. scholarship from Samsung and the John and Claire Bertucci
Graduate Fellowship. Gennady Pekhimenko is supported in part by a Microsoft
Research Fellowship.

\bibliographystyle{ieee}
{\small
\bibliography{ref}
}

\end{document}